\documentclass[10pt,letterpaper]{article}
\usepackage[top=0.85in,left=1.25in,footskip=0.75in,marginparwidth=2in]{geometry}

\newenvironment{keywords}
  {\small\noindent\textbf{\textit{Keywords---}}\space\ignorespaces}
  {\par\medskip}

\usepackage[utf8]{inputenc}
\usepackage[most]{tcolorbox}

\usepackage[round, authoryear]{natbib} 

\usepackage{hyperref}
\usepackage{amssymb}
\usepackage{amsmath}
\usepackage{array}
\usepackage{booktabs}
\usepackage{multirow}
\usepackage{nameref}
\usepackage{tabularx}

\usepackage[right]{lineno}

\usepackage{microtype}
\DisableLigatures[f]{encoding = *, family = * }

\raggedright
\usepackage{changepage}

\usepackage[aboveskip=1pt,labelfont=bf,labelsep=period,singlelinecheck=off]{caption}

\makeatletter
\renewcommand{\@biblabel}[1]{\quad#1.}
\makeatother

\usepackage{lastpage,fancyhdr,graphicx}
\usepackage{epstopdf}
\fancyheadoffset[L]{2.25in}
\fancyfootoffset[L]{2.25in}

\usepackage{color}

\definecolor{Gray}{gray}{.25}

\usepackage{graphicx}

\usepackage{sidecap}
\usepackage{wrapfig}
\usepackage[pscoord]{eso-pic}
\usepackage[fulladjust]{marginnote}
\reversemarginpar

\usepackage{booktabs}

\usepackage{tcolorbox}
\usepackage{enumitem}
\usepackage{lato}
\usepackage{microtype}
\usepackage{needspace}  

\newtcolorbox{researchbox}{
  colback=gray!3!white,
  colframe=gray!60!black,
  fonttitle=\bfseries\small,
  title=\textcolor{white}{\small Box 1. Two fundamental ways to think about replicability and heterogeneity},
  boxrule=0.45mm,
  arc=1.5mm,
  coltitle=black,
  colbacktitle=gray!60!black,
  attach boxed title to top center={xshift=0mm, yshift=-2.5mm},
  boxed title style={size=small, colframe=gray!60!black, sharp corners},
  left=2.5mm,
  right=2.5mm,
  top=2.5mm,
  bottom=2.5mm,
  width=\dimexpr\textwidth-5mm\relax,
  fontupper=\small\fontfamily{lato}\selectfont\linespread{0.95}\selectfont,
  enhanced,
  nobeforeafter
}

\newtcolorbox{researchbox2}{
  colback=gray!3!white,
  colframe=gray!60!black,
  fonttitle=\bfseries\small,
  title=\textcolor{white}{\small Box 2. Is the variance floor specific to Betabinomial?},
  boxrule=0.45mm,
  arc=1.5mm,
  coltitle=black,
  colbacktitle=gray!60!black,
  attach boxed title to top center={xshift=0mm, yshift=-2.5mm},
  boxed title style={size=small, colframe=gray!60!black, sharp corners},
  left=2.5mm,
  right=2.5mm,
  top=2.5mm,
  bottom=2.5mm,
  width=\dimexpr\textwidth-5mm\relax,
  fontupper=\small\fontfamily{lato}\selectfont\linespread{0.95}\selectfont,
  enhanced,
  nobeforeafter
}

\newtcolorbox{researchbox3}{
  colback=gray!3!white,
  colframe=gray!60!black,
  fonttitle=\bfseries\small,
  title=\textcolor{white}{\small Box 3. How many replications do you really have?},
  boxrule=0.45mm,
  arc=1.5mm,
  coltitle=black,
  colbacktitle=gray!60!black,
  attach boxed title to top center={xshift=0mm, yshift=-2.5mm},
  boxed title style={size=small, colframe=gray!60!black, sharp corners},
  left=2.5mm,
  right=2.5mm,
  top=2.5mm,
  bottom=2.5mm,
  width=\dimexpr\textwidth-5mm\relax,
  fontupper=\small\fontfamily{lato}\selectfont\linespread{0.95}\selectfont,
  enhanced,
  nobeforeafter
}

\usepackage{titlesec}
\titleformat*{\section}{\normalfont\large\bfseries}
\titleformat*{\subsection}{\normalfont\normalsize\bfseries}

\begin{document}
\setlength{\textfloatsep}{8pt plus 2pt minus 2pt}
\setlength{\intextsep}{8pt plus 2pt minus 2pt}
\setlength{\floatsep}{8pt plus 2pt minus 2pt}
\vspace*{0.35in}

\begin{flushleft}
{\Large \textbf\newline{The Difference Between ``Replicable'' and ``Not Replicable'' Is Not Itself \textit{Scientifically Replicable}}}
\newline
\\
Erkan O. Buzbas \textsuperscript{1*\ddag},
Berna Devezer \textsuperscript{1,2,3*}
\\

\bigskip
\bf{1} Department of Mathematics and Statistical Science, University of Idaho
\\
\bf{2} Department of Business, University of Idaho
\\
\bf{3} Institute for Modeling Collaboration and Innovation, University of Idaho
\\

\bigskip
* These authors contributed equally to this work.\\
\ddag Corresponding author (erkanb@uidaho.edu)

\end{flushleft}

\vspace{1cm}

\begin{abstract}
Replication studies estimate the replicability rate of scientific results by aggregating binary verdicts of experiments as ``replicated'' and ``not replicated''. These estimates assume rarely attainable exact replications. Experiments often do not share a common data-generating process, leading to non-exact replications. We formalize two statistical interpretations of non-exactness. In a shared latent rate model (benchmark), experiments are exchangeable conditional on a common random replicability rate. In a conditionally independent rates model (operational), each experiment has its own random replicability rate, drawn independently from a population. Under the benchmark model, even small variability in assumed replicability rates induces a variance floor of the estimated mean replicability rate that is irreducible by more replications. Under the operational model, the degree of non-exactness is not identifiable from binary verdict data, which contain no information about between-experiment heterogeneity. Researchers cannot tell which precision regime they are in, and high- and low-replicability sequences cannot be distinguished in principle. Binary verdicts systematically understate uncertainty in reporting scientific results, making high- and low-replicability sequences appear discriminable, when they are not. Common sources of heterogeneity amplify these problems as shown in a reanalysis of Many Labs~4 data. The practice of aggregating replicability rates across heterogeneous literatures produces averages that conflate incommensurable experimental regimes and lack a stable scientific interpretation. Hence, replicability rates as currently used do not provide a reliable demarcation criterion for scientific results. The replication crisis, if there is one, cannot be established by methods used to declare it. 
\end{abstract}


\section*{Introduction}
Science is said to progress through replication. A result that holds up is viewed as trustworthy; one that does not is suspect. Yet how we determine what holds up, the question on which all such assessments rest, is rarely examined. Replicability rates of scientific results are now routinely reported~\citep[e.g.,][]{klein2022many,open2015estimating,camerer2016evaluating}, compared across disciplines~\citep[e.g.,][]{baker2016reproducibility,tyner2026investigating}, and used to pronounce verdicts on entire fields~\citep[e.g.,][]{freedman2015economics,Pashler2012}. The implicit logic is that a high replicability rate signals a healthy literature and a low rate signals a crisis. We argue that this logic does not survive formal statistical scrutiny because the statistical machinery used to convert outcomes of replication experiments into verdicts is inadequate for this task.

An {\em exact replication} of a {\em reference}\footnote{We refer to the original target experiment as \textit{reference} throughout this paper.} experiment is an experiment that differs from the reference only with respect to its randomly and independently generated sample. That is, the design, procedures, measurements, models, and inferential methods of an exact replication experiment are substantively and statistically equivalent to those of the reference experiment. We refer to the inferential target and its associated procedure (e.g., parameter estimation, hypothesis test, model selection) as the result type, and assume it is fixed across the reference and its replications.

In scientific practice, exact replications are the exception rather than the rule. Even carefully designed and well-intentioned replications differ from a reference in ways that may matter statistically. Different subject pools, experimenters, equipment, procedures, times, and places are potential sources of deviations from the reference. Reasons for such undesired deviations may include technical challenges in replicating an experiment exactly under different lab or field conditions, lack of domain knowledge to dictate what conditions are relevant or should be controlled, poor execution of the planned experiment, or unforeseen factors related to data structure, such as missing data. Regardless of the specific reasons, the practical outcome is that most replications are {\em not} statistically exact replications~\citep[see][for example, for an explicit documentation of divergences from the reference and/or challenges in designing exact replications]{errington2021challenges,klein2018many,klein2022many}. We refer to replication experiments that are not statistically equivalent to each other (and hence to the reference) as \emph{non-exact} replication experiments. The present work formalizes the effects of non-exactness and shows how misleading metascientific inferences based on an estimated mean replicability rate can be, even under ideal conditions where the degree of non-exactness is known. 

We study non-exactness under two distinct statistical models, corresponding to two ontological ways non-exactness can arise in practice. Both models speak to what replicability rates can and cannot deliver. However, statistical arguments we advance here are not about familiar observations that replication studies are underpowered or that inferences from the reference experiment and its replications are subject to sampling error. Standard statistical theory is well-equipped to model and account for errors these problems may create. We identify unique inferential challenges that arise from non-exactness: 1) even with impractically large numbers of replications, discriminability between high- and low-replicability results is lost, 2) binary verdicts such as ``this result replicates'' and ``that result does not'' cannot be reliably distinguished from each other using the methods by which they are currently produced, and 3) large-scale replication studies that combine rates across different result types or literatures do not account for the incommensurability of these measures.

Our arguments are inspired by \citet{gelman2006difference}, who showed that the difference between a statistically significant result and a nonsignificant one is not itself statistically significant. Our results may be interpreted as an analogous limitation for replicability. As \citet{gelman2006difference} emphasized, their result was not merely about the arbitrariness of the significance threshold. Even large apparent differences in p-values (such as $p = 0.01$ versus $p = 0.20$) were not themselves statistically significant. Similarly, we are not merely observing that two results with high and low mean replicability rates have an arbitrary boundary where they can be statistically distinguished. We show that two results whose true mean replicability rates are far from any reasonable threshold, such as $80\%$ and $20\%,$ cannot be reliably distinguished from each other, even with impractically large numbers of non-exact replications. Hence, rather than the existence of an arbitrary threshold, the problem is the impossibility of conclusive statements about ``high'' versus ``low'' replicability under realistic conditions. The implication is that replicability rate cannot serve as a reliable demarcation criterion between results that hold and those that do not, and thus, the difference between a replicable result and a not replicable one is not itself scientifically replicable\footnote{This titular claim is not itself a replicability verdict in the sense critiqued in this paper. It refers to a mathematical result about the inferential procedure used to produce such verdicts. The title uses ``scientifically replicable'' colloquially rather than statistically, to mean that the distinction between replicable and not replicable cannot be reliably established by the methods currently used to establish it, not that some specific experiment showing such a difference would fail to replicate.}.

\section*{Statistical theory of replicability rates}

A natural definition of replicability rate of a particular outcome, $r,$ of a fixed result type  given a sequence of random results $R_1, R_2, \cdots, R_m,$ from $m$ exact replications of the reference experiment is
\begin{equation}\label{eq:reprate}
\phi = \lim _{m\rightarrow \infty} \frac{1}{m}\sum_{i=1}^{m}\mathbf{1}_{\{R_i=r\}},
\end{equation}
where $\mathbf{1}_{\{A\}}$ is equal to 1 if $A$ is true, and $0$ otherwise. By this definition, there exists a parameter $\phi \in [0,1]$ of a set of statistically equivalent experiments for a target outcome $r$. The number of replicated results in $m$ independent exact replication experiments conditional on $\phi$ is then a Binomially distributed random variable, with probability of success $\phi.$
Below, we present two statistical models which aim to capture distinct forms of non-exactness in replication experiments. Inferential implications of these models on replicability of scientific studies under the real-life practice are quite different. 

\subsection*{Benchmark model}
Here, we consider a {\em particular random sequence} of replication experiments. Conditional a realization of theory that the sequence embodies, the non-exactness is due to which chosen random sequence realizes, out of the many that the theory admits. The replicability rate governing this particular sequence is not only a fixed property of the result $r$ (as defined by equation~\ref{eq:reprate}), but also a property of the set of replication choices. One way to capture this uncertainty is by treating the replicability rate itself as a random draw from a population, preserving the structure that the Bernoulli draws within the sequence remain conditionally independent given this replicability rate. Marginally, however, the outcomes within a sequence are no longer independent. The key difference from a sequence of exact replications with a fixed replicability rate $\phi$ conditional only on $r,$ is that there are now two sources of uncertainty: the uncertainty due to random sampling within an experiment conditional on the replicability rate and the uncertainty due to randomness of the replicability rate this specific sequence is based on. The first source of variability shrinks with number of replications but the second does not. 

The setup involving a particular random sequence invokes the idea that the binary outcomes from $m$ replication experiments are exchangeable, but not necessarily independent. By de Finetti's representation theorem, any infinitely exchangeable sequence of binary random variables admits a mixture representation, and we assume throughout that finite sequences are embedded in such an infinitely exchangeable sequence. Then, there exists a latent parameter $\phi \in [0,1]$ such that, conditional on $\phi,$ the outcomes from experiments are independently and identically distributed as Bernoulli with success parameter $\phi.$ The conditional distribution of the total number of replicated results is then $X | \phi, m \sim \mathrm {Bin}(m, \phi).$ Treating $\phi|\mu, \rho \sim \mathrm {Beta}(\mu,\rho),$ where $\mu \in [0,1]$ is the mean and $\rho \in [0,1]$ is the intraclass correlation parameter, yields the distribution of the number of replicated results as Betabinomial, with probability mass function
\begin{equation}\label{eq:Betabinomial}
p(x | m, \mu, \rho) = \int_0^1 \mathrm {Bin}(x|m,\phi)\,
\mathrm {Beta}(\phi|\mu,\rho)\,d\phi,
\end{equation}
which we call the \textit{benchmark model} for a replication sequence. Under this model, all $m$ outcomes depend on the same latent parameter $\phi,$ and the intraclass correlation between any two outcomes is $\rho.$ 

The uncertainty due to randomness of the replicability rate of specific sequence can be contextualized through the underdetermination of experiments by theory. A theory rarely specifies all empirical conditions required to instantiate it in an experiment. Sample populations, stimuli, measurement instruments, and procedural details must be filled in by the experimenter, and the theory is typically compatible with many such choices. Each valid instantiation carries its own replicability rate. A replication sequence under the benchmark model realizes one particular set of these choices, so the rate governing the sequence reflects the instantiation that was realized rather than a unique property of the theory. Under this reading, the parameters $\mu$ and $\rho$ acquire a specific interpretation, where $\mu$ is the mean replicability rate across the instantiations the theory admits, and $\rho$ quantifies the degree to which the theory underdetermines its empirical content. A sharply specified theory constrains its valid instantiations and yields small $\rho,$ while an underspecified theory admits heterogeneous instantiations and yields large $\rho.$\footnote{To the extent that instantiation variability arises from valid realizations of the theory rather than from execution error, $\rho$ can be read as an index of theoretical underspecification. However, sources such as execution error can also contribute to $\rho$ without an underdetermined theory. Realized instantiations do not necessarily have to be valid as theory tests, in this sense.} When a theory does not specify which conditions individuate its proper test, metascientific practice treats distinct data generating processes as replications of a single result, and their distinct rates are absorbed into one replicability rate estimate.

The distinction between $\phi$ and $\mu$ determines the inferential target. $X/m$ is a consistent estimator of $\phi.$ But $\phi$ depends on which valid instantiation the sequence happened to realize. The only meaningful rate-like quantity in the model that is invariant to the realized instantiation is $\mu.$ It is defined at the level of the population of valid instantiations, prior to the realization of any particular sequence, and it is the quantity the estimated rate can track across instantiations. 

The natural estimator of $\mu$ is $\hat{\mu} = X/m$ and $\rho$ governs its variance by
\begin{equation}\label{eq:variancebenchmark}
\mathbb{V}(\hat{\mu} | m, \mu, \rho) = \mu(1-\mu)\left[\frac{1}{m} + \frac{(m-1)}{m}\rho\right].
\end{equation}

As $m \to \infty,$ equation~\ref{eq:variancebenchmark} converges to $\mu(1-\mu)\rho,$ which is positive whenever $\rho > 0$ and $\mu \notin \{0,1\}.$ This quantity is an {\em irreducible variance floor} of $\hat{\mu}$ as an estimator of $\mu.$ Adding replications does not eliminate this floor because the variability captured by $\rho$ reflects the random draw of $\phi$ from $\mathrm{Beta}(\mu, \rho),$ and additional outcomes from the same sequence cannot resolve the population-level uncertainty about $\mu$ that this draw introduces.

Replication practice, meanwhile, reports rates as properties of results by comparing them across labs, literatures, and time, and pronouncing results replicable or not on the basis of those comparisons. If $\phi$ were the target, the statement ``this result replicates 36\% of the time'' would have to be read as ``this particular sequence happened to produce 36\%, and a different sequence could produce something quite different,'' which is not what any replication report intends to convey. The claims of replication practice are therefore claims about $\mu,$ and we take $\mu$ as the target for that reason. 

Replication projects do not follow the benchmark model. No instantiation is identified in advance and carefully sampled; instead, each experiment's instantiation takes shape as it is run, through choices made at that site, so a sequence of $m$ replications settles the unspecified conditions $m$ times rather than once. The benchmark model serves as an insightful, if not realistic, idealization. It is a model in which the consequences of dispersion for estimating $\mu$ can be quantified. Even under this idealization, the design of a replication sequence is unable to deliver desirable inferential properties of $\hat{\mu}$. A more realistic representation would be each non-exact replication experiment contributing a single binary verdict about the target result under \textit{its own replicability rate}~\citep[see][]{buzbas2023}.

\subsection*{Operational model}

Here, we consider a sequence of $m$ non-exact experiments where each experiment has its own replicability rate, drawn independently $\phi_i \sim \mathrm{Beta}(\mu, \rho),\; i=1,2,\cdots, m.$ Within experiment $i,$ with replicability rate $\phi_i,$ the $k_i$ binary replication outcomes $(R_{i1}, R_{i2}, \cdots, R_{ik_i})$ are conditionally independent, and $k_i$ track exact replications within the instantiation that experiment $i$ realizes. So, each experiment potentially can produce $k_i > 1$ outcomes under this model if exact replications were possible, but a typical realization in practice is $k_i=1,$ for all $i.$  The number of replicated results from experiment $i$ is $X_i | \phi_i, k_i \sim \mathrm {Bin}(k_i, \phi_i).$ With $\phi_i \sim \mathrm {Beta}(\mu, \rho),$ independent across experiments from the mixing distribution, each $X_i$ marginally follows a Betabinomial distribution with probability mass function
\begin{equation}\label{eq:MarginalBetabinomial}
p(x_i | k_i,\mu,\rho) = \int_0^1 \mathrm{Bin}(x_i | k_i,\phi)\, \mathrm{Beta}(\phi | \mu,\rho)\,d\phi,
\end{equation}
which we call the \textit{operational model} for a replication sequence. 

The natural estimator of $\mu$ here is $\hat{\mu} =  (1/m)\sum_{i=1}^m \left(X_i/k_i\right),$ and $\rho$ governs its variance by
$$\mathbb{V}(\hat{\mu} \mid k_1,\cdots, k_m, m, \mu, \rho) = \mu(1-\mu) \left[(1 - \rho)/m^2 \sum_{i=1}^m(1/k_i) + \rho/m \right],$$ 
(see Appendix for details). Two limiting cases bound this variance. 
\begin{itemize}
\item Infinite exact replications: $k_i \to \infty,$ for all $i.$ We have $$\mathbb{V}(\hat{\mu} | m, \mu, \rho)=\frac{\mu(1-\mu)\rho}{m},$$ which is a lower bound on the variance of $\hat{\mu}$ under the operational model for fixed $m.$ Here, Binomial sampling variability vanishes and the remaining variability comes entirely from between-experiment heterogeneity, governed by $\mu,$ $\rho,$ and $m.$ Unlike the benchmark model's irreducible floor, this quantity converges to zero as $m \to \infty.$

\item No exact replications: $k_i = 1,$ for all $i.$ We have
\begin{equation}\label{eq:var3}
\mathbb{V}(\hat{\mu} | k = 1, m, \mu)=\frac{\mu(1-\mu)}{m},
\end{equation}
where $\rho$ drops out. In contrast to the benchmark model, where the $m$ outcomes are exchangeable and correlated, under the operational model when $k_i = 1,$ the Betabinomial distribution for the number of successes collapses to mutually independent Bernoulli variables. The total number of replicated results is $X \sim \mathrm{Bin}(m,\mu)$ with variance $m\mu(1-\mu).$ It is this independence that causes $\rho$ to drop out from the variance in equation~\ref{eq:var3}. Higher moments, which carry information about $\rho,$ require $k_i \geq 2.$  When only one binary replication outcome is observed per experiment, $\rho$ is not an identifiable parameter from the data and hence the theoretical bound on what we can learn from non-exact replications when their results are reduced to binary verdicts. No amount of non-exact replications will ever let us learn about $\rho$ because the data do not contain that information.
\end{itemize}

The underdetermination reading of the benchmark model extends to here, with one caveat. In the operational model with $k_i = 1$, no instantiation is realized twice. Each experiment fills in the empirical conditions the theory leaves open with its own choices of stimuli, instruments, and procedure, and thereby realizes its own instantiation, with its own rate $\phi_i.$ The population $\mathrm{Beta}(\mu, \rho)$ is the same in the benchmark and operational models, and represents the set of valid instantiations of the theory with their replicability rates. The models differ in how this population is sampled in a sequence of replications. The benchmark model samples the instantiation space once and observes $m$ outcomes from the draw and the operational model samples it $m$ times and observes one outcome per draw. The mean replicability rate, $\mu,$ retains the same meaning in both models, but $\rho$ does not. 

The operational model is a faithful description of the replication practice. Replication projects do not fix an instantiation and replicate within it. Instead, each lab fills in the unspecified conditions independently. This practice comes at an inferential cost, because under the operational model the dispersion parameter leaves no trace in the data. The model describes the practice while being unable to estimate the properties of the replication rate. We therefore use the operational model to characterize what data from replication practice can and cannot inform, and the benchmark model to characterize what the information loss entails inferentially at a given degree of non-exactness. 

\subsection*{Interpretation of $\rho$ and bounds on the distribution of $\hat{\mu}$}
Mathematically, the benchmark and operational models have 
 $$\rho = \mathbb{V}(\phi)/[\mu(1-\mu)],$$ 
where $\rho$ measures how replicability rates vary across instantiations the theory admits. However, the two models assign different roles to $\rho.$ Under the benchmark model, it is the intraclass correlation of a shared exchangeable sequence. Under the operational model, it is the normalized variance of the distribution from which independent experiment-specific replicability rates $\phi_i$ are drawn. What remains to be characterized is when this structural difference is observable. At $m=1,$ the benchmark and operational models collapse to a single $\mathrm{Betabinomial}$ draw, because the structural feature distinguishing them disappears. At $m > 1,\;k = 1,$ the operational model's $m$ independent $\phi_i$'s induce $m$ marginally independent Bernoulli outcomes, and with no within-experiment replication, it is not possible to detect the effect of $\rho$ on the variability of the mean replicability rate. In these cases, the information content of the data cannot resolve structural differences between models. The substantive contrast between exchangeability under the benchmark model and independence under the operational model is empirically observable only when $m > 1$ and $k > 1.$ Typical replication practice satisfies only $m > 1.$

Distinct meanings of $\rho$ under the two models have inferential implications on the mean replicability rate $\mu.$ Under the benchmark model, estimates of the mean replicability rate remain highly variable regardless of the number of replications. Under the operational model at $k = 1,$ which most closely represents typical replication practice, the variability of the estimated mean replicability rate is the same regardless of the degree of non-exactness. A researcher cannot determine from binary verdicts alone whether the effect of $\rho$ is negligible or severe. 

Box~1 summarizes the two models and their interpretations in non-technical terms. In the remainder of this section, we discuss the bounds on the distribution of $\hat{\mu}$ as a function of $\rho.$ We start with the ideal case of exact replications.
\begin{figure}[p]
\begin{researchbox}
\vspace{0.3 cm}

\noindent
{\bf 1. Benchmark model: What would variability look like if we knew the heterogeneity between experiments?}
Imagine a sequence of replication experiments targeting the same result. Some experiments are easy to replicate, some are hard, and suppose we knew how much they differ from each other.  We think about
\begin{itemize}
    \item $\mu$: The mean replicability rate across possible replication sequences. It answers: ``On average, what fraction of experiments in a replication sequence replicates a target result?''
    \item $\rho$: How variable the underlying replicability rate is across possible replication sequences. It answers: ``How much does non-exactness inflate the variability of the estimated replicability rate beyond sampling variability?''
\end{itemize} 
Under the benchmark model (equation~\ref{eq:Betabinomial}), $\hat{\mu} = X/m$ estimates the mean replicability rate from $m$ replications. When $\rho$ is small, replication outcomes behave nearly independently and $\hat{\mu}$ concentrates tightly around $\mu.$ When $\rho$ is large, outcomes are strongly correlated, $\hat{\mu}$ is more variable, and distinguishing high- from low-replicability regimes grows difficult even with many replications. This model answers the hypothetical question ``if the degree of non-exactness $\rho$ were known, how much would that heterogeneity alone inflate the variability of the estimated replicability rate and blur the distinction between \textit{good} and \textit{bad} results?'' This is what Figure~\ref{fig:variability.phi} displays.
\vspace{0.3 cm}

\noindent
{\bf 2. Operational model: What can we actually learn from one binary verdict per experiment?}
In real replication projects, each experiment is often replicated only once and its result is reported as a single binary verdict: {\em replicated} or {\em not replicated.} Under the operational model, all $k = 1$, the $m$ verdicts are mutually independent, each with success probability $\mu.$ The consequence is that independence replaces exchangeability, and $\rho$ drops out from the variance of the mean replicability rate. Sequences generated under very different heterogeneity patterns (different $\rho$) can produce the same pattern of successes and failures, if they share the same $\mu$.
\vspace{0.3 cm}

\noindent
{\bf Why this matters for replication experiments.}
An analogy may help. The benchmark model is like taking $m$ temperature readings from the same thermometer. The readings are correlated because they depend on the same instrument. Averaging many readings pins down what this thermometer says, but not whether this thermometer runs hot or cold. This calibration error is a property of the instrument we picked, and no number of readings from the same instrument reveals it. The operational model at $k \to \infty$ is like reading each of $m$ different thermometers infinitely many times, so that each instrument's own value is known exactly. The instruments still disagree with each other, but the average across many independent instruments converges to the true mean temperature as $m$ grows. But the operational model at $k = 1$ is like reading each of $m$ different thermometers exactly once. Each read gives us an estimate, but we cannot tell whether the thermometers are well-calibrated or wildly miscalibrated in different directions. The readings look the same either way.
\vspace{0.3 cm}

The benchmark model (equation~\eqref{eq:Betabinomial} and Figure~\ref{fig:variability.phi}) shows that, \emph{if} experiments differ in replicability rate, heterogeneity increases variability in the estimated mean replicability rate $\hat{\mu},$ which in turn makes high- and low-replicability challenging to discriminate, even when the degree of non-exactness $\rho$ is known. The operational model with all $k = 1$ shows that typical replication data, which involves one binary verdict per experiment, cannot recover that heterogeneity at all. Different heterogeneity regimes look identical if $\mu$ is the same. Together, these two models frame our core claim. Heterogeneity in true replicability rate impacts the diagnostic value of replicability rates but cannot be detected within typical replication practice.

\end{researchbox}
\end{figure}
\vspace{0.2cm}

\begin{figure}[h] 
\begin{center}
\includegraphics[width=\textwidth]{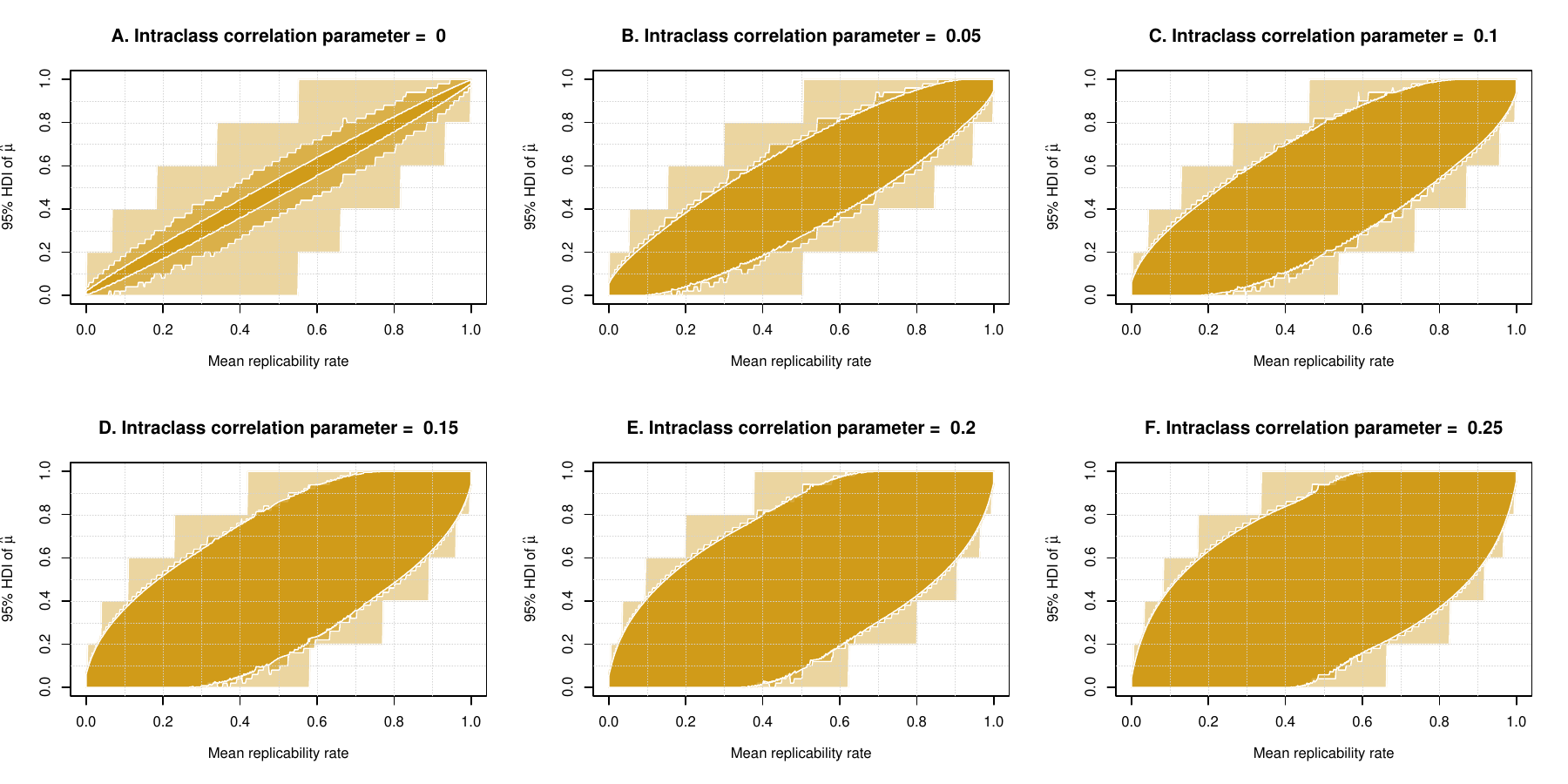}
\end{center}
\caption{Theoretical sensitivity of the estimated mean replicability rate $\hat{\mu}$ to the degree of non-exactness $\rho,$ under the benchmark model (equation~\ref{eq:Betabinomial}). Each panel plots the $95\%$ HDI of the sampling distribution of $\hat{\mu} = X/m$ as a function of $\mu,$ for $m \in\{5, 50, 500\}$ replications (light to dark). Panel A: exact replication ($\rho = 0$). Panels B--F: fixed non-exactness regimes $\rho \in\{ 0.05, 0.10, 0.15, 0.20, 0.25\}$. These panels do not represent what is estimable from binary replication verdicts alone, where $\rho$ is not identifiable (see operational model). They show the discriminability consequences of a given $\rho$ when that value is determined by an external source of heterogeneity, as in the numerical examples and tables that follow.}
\label{fig:variability.phi}
\end{figure}

\noindent
{\bf Sequence of exact replications.} If $m$ experiments are exact replications of each other, then $\mu=\phi_1=\phi_2=\cdots=\phi_m,$ and $\rho=0.$ This is the theoretical ideal from which replication practice departs. We have $X \sim \mathrm {Bin}(m,\phi),$ and $\mathbb{V}(\hat{\mu}|\mu,m)=\mu(1-\mu)/m.$ As $m$ increases, the variance of $\hat{\mu}$ decreases at rate $1/m,$ and $\hat{\mu}$ is a consistent estimator of $\mu$. Panel A of Figure~\ref{fig:variability.phi}, computed under this model, gives a sense of the convergence. At $\rho = 0$ the benchmark and operational models coincide, so panel A applies to both. The mean replicability rate of an experiment (horizontal axis) is plotted against the 95\% Highest Density Interval (HDI) of the sampling distribution of $\hat{\mu}$ (vertical axis). The shades from light to dark represent the number of replications $m = 5,\; 50,$ and $500,$ respectively. At smaller number of replications, estimates of mean replicability rates vary greatly, even for exact replications.

As a numerical illustration of discriminability of high- and 
low-replicability cases under exact replications, we take the true mean replicability rate as $\mu_H= 0.80,$ for the high-replicability case. The 95\% HDI of the sampling distribution of $\hat{\mu}_H$ based on $m=50$ replications is $(0.70, 0.90).$ For the low replicability rate case, we take the true mean replicability rate as $\mu_L=0.20$. The 95\% HDI of the sampling distribution of $\hat{\mu}_L$ based on $m=50$ replications is $(0.10, 0.30).$ The two intervals do not overlap, and hence the high- and low-replicability rate results are discriminable theoretically at $0.95$ probability level for $m=50$ replication experiments. We say \textit{theoretically}, because the comparison is made with true probabilities of the sampling distribution of $\hat{\mu}.$ They are not estimates based on realized data. In practice, the sampling distribution of $\hat{\mu}$ is unknown and these probabilities have to be estimated by building interval estimates for $\mu.$ Consequently, there will be additional uncertainty in assessing the difference between high- and low-replicability results from a given sequence of replication experiments.

It is clear from Figure~\ref{fig:variability.phi} panel A that the discriminability of high- and low-replicability results for exact replications depends on what the desired level of comparison is and how many replications are performed. If a lower bar is set for high-replicability and/or a higher bar for low-replicability, or fewer replications are performed, the 95\% HDIs of the sampling distribution of $\hat{\mu}$ will have more overlap.

\vspace{0.2cm}

\noindent
{\bf Sequence of non-exact replications.} Under the operational model at $k = 1,$ the number of replicated results follows $\mathrm {Bin}(m, \mu)$ regardless of the degree of non-exactness, and a researcher computing an interval estimate for $\mu$ from binary verdicts will obtain intervals that look like those of panel A. As established above, the data cannot reveal which non-exactness regime actually applies. A researcher reporting panel A precision may be correct, or may be severely underestimating uncertainty. Either way, the data will not provide a way to tell which.

The intervals displayed in panels B--F and the discriminability failures they reveal are properties of the benchmark model with known $\rho>0.$ Here the picture changes fundamentally, and not in a direction that more replications can fix. Under the benchmark model, the variance of $\hat{\mu}$ (equation~\ref{eq:variancebenchmark}) contains the term $[(m-1)/m]\rho,$ which converges to $\rho$ as $m$ grows, and the variance asymptotically approaches $\mu(1-\mu)\rho.$ This causes the sampling distributions of $\hat{\mu}$ for high- and low-$\mu$ to overlap considerably for realistically large $m$. Adding more replications decreases the variance initially, but ultimately it hits the irreducible floor that $\rho$ itself determines\footnote{Throughout, we treat $\mu$ and $\rho$ as fixed parameters of a given replication sequence; that is, as properties of the data-generating process rather than functions of $m.$ In practice, expanding a replication program may change the stochastic process of the sequence and therefore $(\mu, \rho)$ themselves. This can move the floor in either direction. Including more heterogeneous replications raises $\rho$ and hence, the floor, while including more homogeneous replications lowers $\rho$ and the variance floor with it. Theoretical results here characterize any fixed sequence with $\rho > 0$ and do not depend on which direction expansion takes. The direction most relevant to practice is upward. Replication programs that expand to recruit diverse sites, increasingly the norm in large-scale multi-lab projects, tend to increase $\rho$ beyond what the fixed-sequence analysis suggests, for a given $m.$}. This is the floor on the variance of $\hat{\mu}$ introduced by the random draw of $\phi$ from $\mathrm{Beta}(\mu, \rho).$ Estimation of $\phi$ (as opposed to $\mu$) would be unaffected by this floor, but $\phi$ is not the inferential target of replication practice. Models with more assumptions than the Betabinomial may be able to pool across experiments and may recover partial information about $\rho$ when differences between experiments are small, but the core result holds for any non-degenerate model of $\phi$ (see Box~2). 

\begin{figure}[htbp]
\begin{researchbox2}
\vspace{0.3cm}

\noindent
No. Under the benchmark model, where all $m$ outcomes share a single
latent replicability rate $\phi,$ the irreducible variance floor is $\mu(1-\mu)\rho,$ where $\rho = \mathbb{V}(\phi)/[\mu(1-\mu)]$ and $\mathbb{V}(\phi)$ is the variance of the replicability rate across replication sequences. This variance is a property of whatever distribution generates $\phi,$ not of the Beta family specifically. 

\medskip
As an example, suppose that instead of a Beta distribution, the replicability rate takes only two values: $\phi = \mu + \delta$ with probability $1/2$ and $\phi = \mu - \delta$ with probability $1/2,$ for some $\delta > 0.$ Then $\mathbb{E}(\phi) = \mu$ and $\mathbb{V}(\phi) = \delta^2.$ The intraclass correlation is $\rho = \delta^2/[\mu(1-\mu)]$ and the variance floor is $\mu(1-\mu)\rho = \delta^2,$ which is identical in structure to the Betabinomial case. We chose Beta distribution for convenience because it is the natural mixing distribution, but any non-degenerate mixing distribution produces a floor of the same form. The core results of this paper on the irreducible variance floor and the loss of discriminability would hold for any such distribution.

\end{researchbox2}
\end{figure}

Another way to understand the effect of $\rho$ under the benchmark model is through the effective number of independent replications $$m_{e}=\frac{m}{1+(m-1)\rho},$$ which is the number of exact replications that would yield the same variance of $\hat{\mu}$ under a Binomial model. When $\rho = 0,$ $m_e = m$ and every replication is fully informative. As $\rho$ increases, $m_e$ shrinks. The correlation among outcomes means that additional replications carry less new information. In the limit $\rho \to 1,$ $m_e \to 1$ regardless of $m.$ To develop intuition about what this means in a scientific context, see the illustration in Box~3 and Figure~\ref{fig:effective}.

\begin{figure}[h]
\begin{researchbox3}
\vspace{0.3cm}

\noindent
Under the benchmark model, where all $m$ outcomes share a single latent replicability rate $\phi,$ the effective number of replications is $$m_e = \frac{m}{1 + (m-1)\rho},$$ the Binomial equivalent of the number of independent exact replications that would yield the same variance of $\hat{\mu}.$ Under the operational model at $k=1,$ where each experiment has its own independent $\phi_i,$ the effective sample size is $m.$ The examples below ask ``if the benchmark model is the correct description, how informative are these projects?''.

\medskip
{\bf The Reproducibility Project: Psychology} replicated $100$ results from diverse literatures, each by a different lab, and reported a $36\%$ replicability rate~\citep{open2015estimating}. The sequence sampled results from different research areas, effect sizes, methodological traditions, and replication teams. At a hypothetical conservative $\rho = 0.10,$ the $100$ replications carry the evidential weight of $9$ exact ones. At $\rho = 0.20,$ the weight drops to $5.$ The reported $36\%$ rate is therefore not an estimate obtained from $100$ independent data points. It could be an estimate obtained from as few as or fewer than $9$ data points, depending on the true degree of non-exactness.

\medskip
{\bf The SCORE project} replicated $274$ results across multiple disciplines, mostly with one replication per result~\citep{tyner2026investigating}. At $\rho = 0.10,$ the effective sample size is approximately $10.$ Almost tripling the number of replications from $100$ to $274$ increases $m_e$ from $9$ to $10,$ essentially a gain of one effective replication. The diminishing returns imposed by non-exactness are severe and not remediable by scaling up.

\end{researchbox3}
\end{figure}

\begin{figure}[h]
\begin{center}
\includegraphics[width=\textwidth]{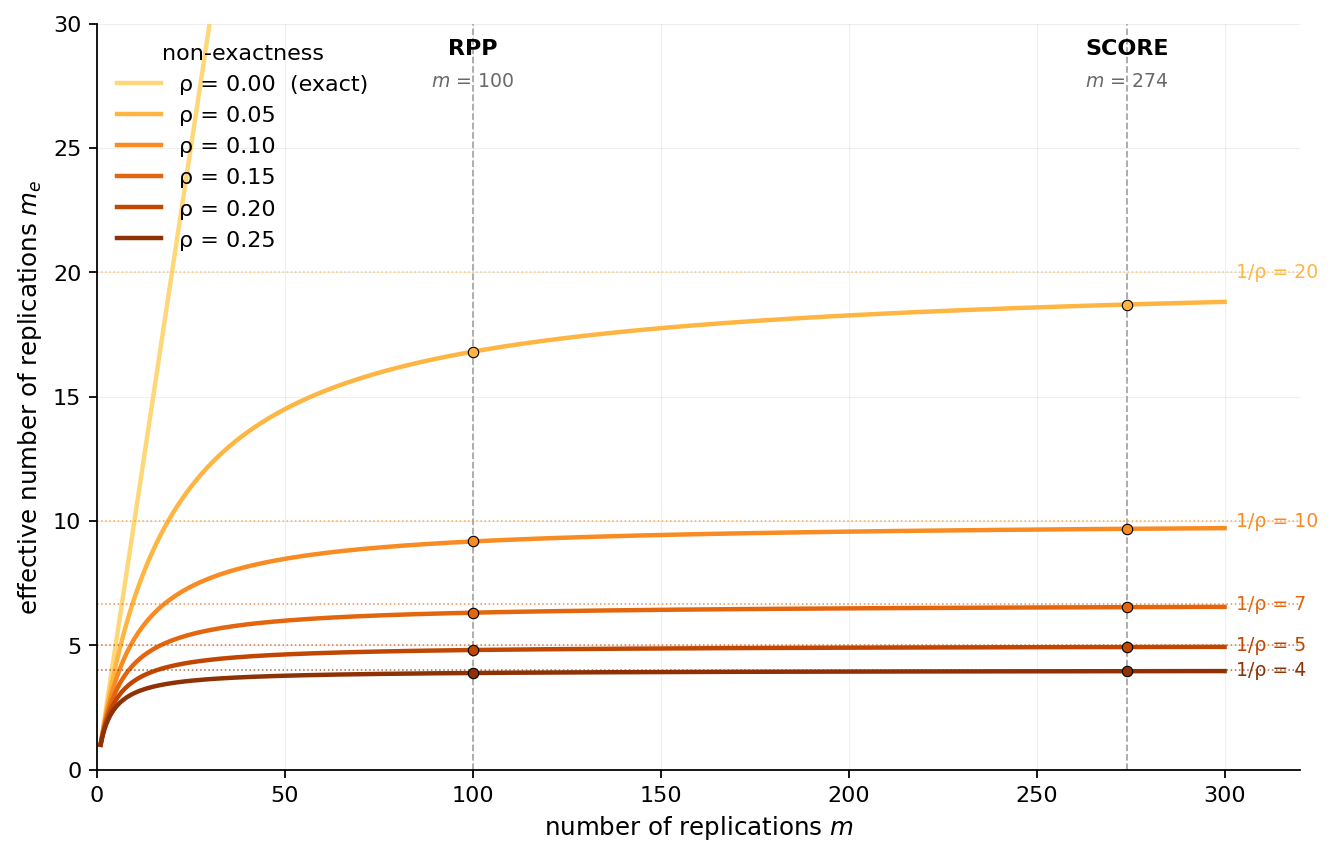}
\end{center}
\caption{Effective number of independent replications $m_e = m/[1+(m-1)\rho]$ as a function of the number of replications $m,$ shown for six values of the intraclass correlation parameter $\rho.$ Horizontal dotted lines mark the asymptote $1/\rho$ for each $\rho > 0.$ Vertical dashed lines locate the Reproducibility Project: Psychology (RPP, $m = 100$) and the SCORE project ($m = 274$). Colored dots mark the implied $m_e$ at each project's replication count for each $\rho$ value. At $\rho = 0,$ the growth is $1:1$ linear with $m,$ reflecting that under exact replications every replication is fully informative.}
\label{fig:effective}
\end{figure}

To show the effect of $\rho$ on the discriminability of replicability rates, panels B--F of Figure~\ref{fig:variability.phi} display the sampling distribution of $\hat{\mu}$ under the benchmark model at six levels of non-exactness: $\rho = 0.05$ (panel B), increasing by increments of $0.05,$ through $\rho = 0.25$ (panel F). Using these plots, we develop intuition about three related properties of the distribution of $\hat{\mu}$ obtained from a sequence of non-exact replications. The first property is about the increase in the width of probability intervals as the degree of non-exactness increases. In contrast to panel A where replications are exact, we intuitively expect an increase with $\rho$. This is illustrated by a comparative evaluation of intervals of the same shade across panels B-F. The second property is about the slow decrease of the width of probability intervals as the number of replications increases. Again in contrast to panel A, where the width of probability intervals shrinks rapidly as the number of exact replications increases, in panels B-F the width of probability intervals decreases slowly even for small values of $\rho.$ Finally, we notice that the width of HDIs is larger for moderate (vs. extreme) levels of mean replicability rate. As such, a replication sequence that replicates, say, $60\%$ of the time (i.e., moderate replicability rate) will be increasingly difficult to distinguish from sequences with both high- and low-replicability rates, as $\rho$ increases or $m$ decreases.

To contrast non-exact replications with exact replications, we again take the number of replications $m=50,$ and the true mean replicability rate of the high-replicability result as $\mu_H = 0.80.$ 95\% HDI of the sampling distribution of $\hat{\mu}_H$ when $\rho=0.15$ (panel D) is $(0.46, 1.00).$ For its counterpart, the low-replicability result with true mean replicability rate $\mu_L=0.20,$ the 95\% HDI of the sampling distribution of $\hat{\mu}_L$ is $(0,0.54).$ The intervals overlap, and hence high- and low-replicable results are {\em not} discriminable theoretically at $0.95$ probability level. To illustrate what inferential challenges look like in practice, in the next section we consider joint inference on $(\mu, \rho).$

\vspace{0.2cm}

\subsection*{Inference under non-exact replications}

Under the benchmark model $(\mu,\rho)$ can in principle be estimated jointly from the aggregate count $X$ of replicated results. If the sampling distribution with known parameters already resists discriminability, however, estimation from realized data cannot fare better. Indeed, even in an unrealistically favorable scenario, where a sequence of $m=100$ replications is available {\em and} the observed and expected number of replicated results are equal, $x=m\mu,$ the marginal posterior distributions of $\mu$ for different true rates overlap so extensively that clean discrimination between high- and low-replicability is practically impossible. We illustrate this effect by investigating the behavior of the marginal posterior distribution of $\mu$ under two prior specifications. The Standard Uniform prior on $\rho$ assigns equal probability to all degrees of overdispersion. The Jeffreys prior concentrates mass near the boundaries of the parameter space, providing a contrasting reference specification.
 
Given $x$ observed replicated results in $m$ experiments, the joint posterior distribution of $(\mu,\rho)$ is 
$$p(\mu, \rho|x, m) \propto p(x|\mu, \rho, m)p(\mu)p(\rho),$$ 
where the likelihood is Betabinomial and independent priors on $\mu$ and $\rho$ are assumed. The interest is in $\mu,$ hence the marginal posterior distribution $$p(\mu|x, m) = \int_{0}^{1}p(\mu, \rho|x, m)d\rho.$$ 

\begin{figure}[h] 
\begin{center}
\includegraphics[width=\textwidth]{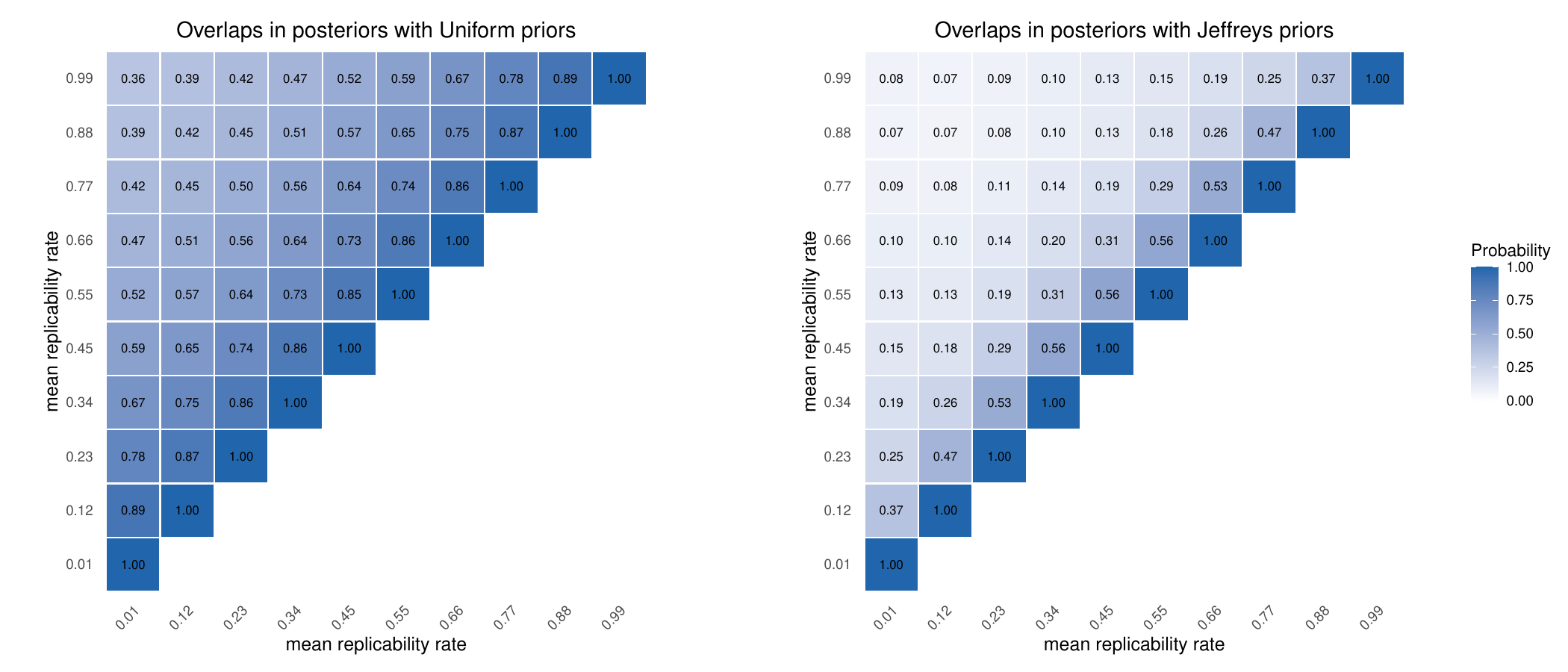}
\end{center}
\caption{Pairwise probability mass overlap between marginal posterior distributions of $\mu$ under Standard Uniform (left) and Jeffreys (right) priors on $\mu$ and $\rho$. Each cell shows the overlap between $p(\mu| x = m\mu_i, m)$ and $p(\mu| x = m\mu_j, m)$ for $\mu_i,\mu_j \in \{0.01, 0.12, \dots, 0.99\}$. The number of replications is $m = 100.$ The overlaps across the grid indicate that even with this highly favorable design, the data provide limited information on discriminating mean replicability rates.}
\label{fig:difference.distinguishability}
\end{figure}

Figure~\ref{fig:difference.distinguishability} shows the posterior probability overlap of pairs of $\mu$ values under the Standard Uniform prior (left) and Jeffreys prior (right).

\begin{figure}[h] 
\begin{center}
\includegraphics[width=\textwidth]{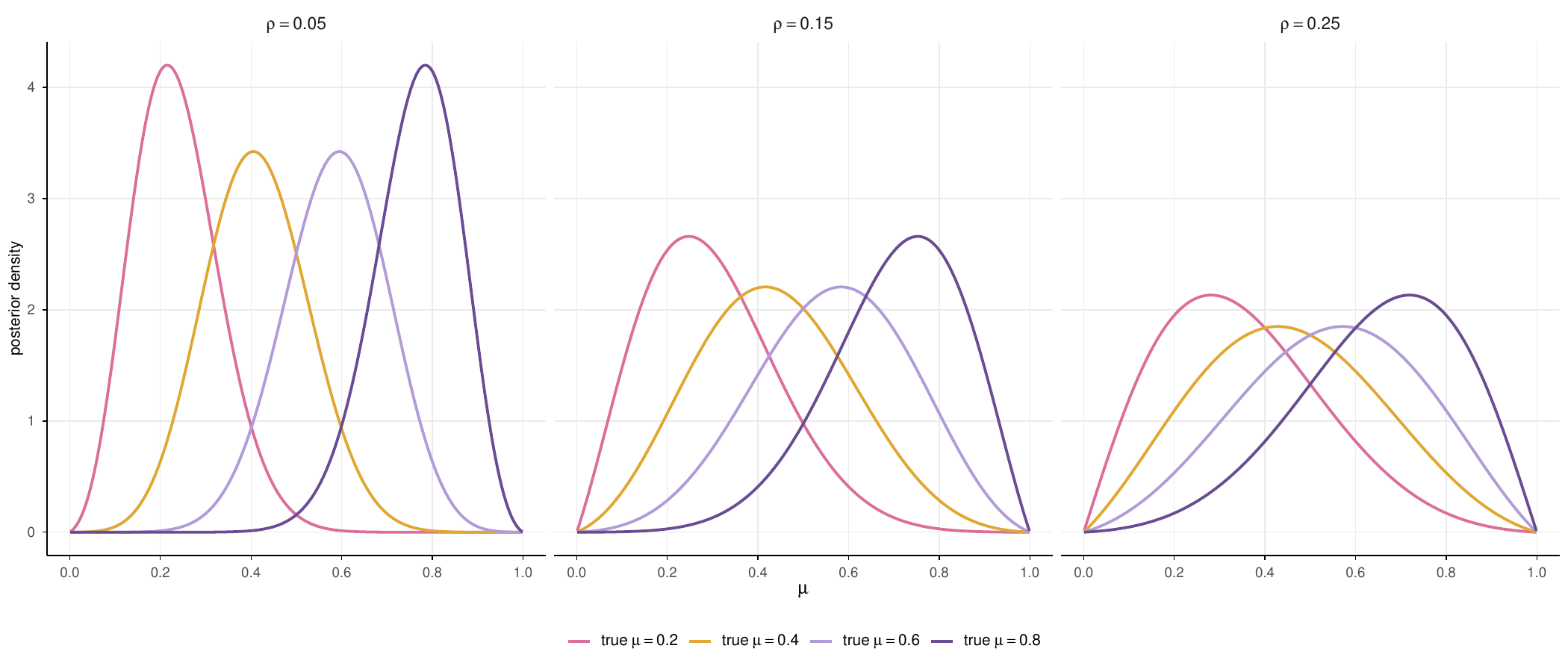}
\end{center}
\caption{Conditional posterior distributions of $\mu$ with $m = 100$ replications and observed count $x = m\mu$, under Uniform prior on $\mu$ and three fixed values of the intraclass correlation parameter: $\rho = 0.05,$ $0.15,$ and $0.25.$ Each panel shows four posteriors corresponding to four data-generating values of $\mu \in \{0.2, 0.4, 0.6, 0.8\}$ when $\rho$ is {\em known}. Assuming $\rho = 0.05,$ the posteriors for $\mu = 0.2$ and $\mu = 0.8$ are discriminable. As $\rho$ increases, the posteriors flatten and overlap progressively, and by $\rho = 0.25,$ the posteriors become practically indistinguishable for most $\mu$ values.}
\label{fig:conditional}
\end{figure}

The posterior overlaps are much larger under the Standard Uniform than under Jeffreys. This means that the data (via the Betabinomial likelihood) on their own do relatively little to separate different values of $\mu.$ Under the Standard Uniform, the smallest overlap is about $0.36$, occurring for the most extreme pair of $\mu$ values $(0.99, 0.01)$. In other words, even when the true mean replicability rates between high- and low-replicability cases differ as much as they possibly can, the marginal posteriors for $\mu$ still retain more than one-third of their probability mass in common and cannot be discriminated from one another. Under Jeffreys prior the minimum overlap drops to about $0.07,$ but this reflects Jeffreys favoring the extreme values of parameters rather than sharply informative replication data. The shift from substantial overlap under the Standard Uniform to little overlap under Jeffreys shows that the apparent discriminability is driven by the prior itself and not the likelihood. Any realistic dataset with sampling variability would likely increase the overlaps and further impede discriminability.

Figure~\ref{fig:conditional} unpacks the marginals of Figure~\ref{fig:difference.distinguishability} under Uniform prior by conditioning on select values of $\rho.$ That is, $\rho$ is assumed {\em known.} Even at $\rho=0.05,$ there is a large overlap between posterior distributions of moderately different replicability rates, making them hard to tell apart. As $\rho$ increases, the data have difficulty discriminating even largely different true replicability rates. In practice, $\rho$ is {\em unknown} and not identifiable from binary verdict data. The marginal posteriors used Figure~\ref{fig:difference.distinguishability} average over conditional posteriors. The researcher inherits the overlap of the least favorable regimes without being able to rule them out.

In the following section, we explore two numerical examples that illustrate the effect of common unsystematic and systematic sources of non-exactness in replications on discriminability of high- and low-replicability under concrete generative models of between-experiment heterogeneity.

\section*{Numerical examples}
Both examples in this section make additional modeling assumptions beyond the Betabinomial model and do not reduce experimental outcomes to binary replication verdicts. Rather, the parameters of the full model are used and then mapped to mean replicability rate $\mu$ and intraclass correlation $\rho.$ Each experiment or lab contributes $k = 1$ binary outcome, as is typical in replication practice. The large-$n$ results in the examples use large-$n$ approximations, which treat within-experiment variability as negligible. Results from the examples are not exercises in estimation. They calculate $\mu$ and $\rho$ from a fully specified generative model rather than estimating them from a sequence of binary outcomes. The goal is to show what implied $\rho$ values would be when concrete physical or statistical sources of non-exactness are present, and map those implied values to the panels of Figure~\ref{fig:variability.phi}. Once those implied $\rho$ values are in hand, the inferential limitations established earlier apply directly. They determine how wide the sampling distribution of $\hat{\mu}$ is and how poor discriminability becomes under realistic experimental conditions.

\subsection*{1. Effect of sample heterogeneity on replicability rate}
The first source of non-exactness we examine is heterogeneity in the population from which each replication draws. The population of interest could be people, cell lines, soil samples, firms, or ecological sites. What matters is that replication experiments draw samples from populations whose true effects differ from each other, even when every other aspect of the protocol is held constant. In human behavioral sciences, sample composition is a well-recognized instance of this~\citep{henrich2010most,bryan2021behavioural}, but the structure of the problem is general. Here we ask what population heterogeneity implies for the replicability rate.

Each of the $i = 1, 2, \ldots, m$ replication experiments draws a sample of $n$ subjects from a population with true effect
$$\theta_i = \theta + \varepsilon_i, \quad \varepsilon_i \sim \mathrm {Nor}(0, \sigma^2).$$ Here the variance $\sigma^2$ captures how much the true effect in each replication experiment varies due to its subject pool. The overall scientific interest is in the sign of $\theta,$ the common underlying effect. Experiment $i$ tests the local hypothesis $H_{0i}: \theta_i \leq 0$ using the estimated effect size $\hat{\theta}_i \sim \mathrm {Nor}(\theta_i, \mathrm{SE}^2),$ where $\mathrm{SE} = f/\sqrt{n},$ $\sigma_s$ is the within-population standard deviation, and $n$ is the sample size. Under non-exactness, $\theta_i$ varies across experiments, so different experiments are testing different hypotheses. We count $1$ from experiment $i$ if $\hat{\theta}_i > 0,$ that is, if the estimated effect is positive. The replicability rate for experiment $i$ is
\begin{equation*}
\phi_i = P(\hat{\theta}_i > 0 \mid \theta_i, \mathrm{SE}) =
\Phi\!\left(\frac{\theta_i}{\mathrm{SE}}\right) =
\Phi\!\left(\frac{\theta + \varepsilon_i}{\mathrm{SE}}\right).
\end{equation*}
The mean replicability rate across experiments is
\begin{equation}\label{eq:phi.ex1}
\mu = \Phi\!\left(\frac{\theta}{\sqrt{\mathrm{SE}^2 + \sigma^2}}\right),
\end{equation}
(see Appendix for details). The mean replicability rate $\mu$ combines two sources of variability: $\mathrm{SE}$ due to within-experiment sampling variability and $\sigma$ due to between-experiment heterogeneity. As either source of variability grows, $\mu$ goes to $0.5.$ The intraclass correlation parameter $\rho$ is 
\begin{equation*}
\rho =\frac{\mathbb{V}(\phi_i)}{\mu(1-\mu)}=\frac{\mathbb{E}[\phi_i^2] - \mu^2}{\mu(1-\mu)} =  \frac{\mathbb{E}\!\left[\Phi\!\left(\frac{\theta +
\varepsilon_i}{\mathrm{SE}}\right)^2\right] - \mu^2}{\mu(1-\mu)}.
\end{equation*}

In the large-$n$ limit, $\mathrm{SE} \to 0$ and equation~\ref{eq:phi.ex1} reduces to $\mu = \Phi(\theta / \sigma),$ approximating the power of the common test when within-experiment sampling variability is negligible relative to between-lab heterogeneity. For large but finite $n,$ $\mathrm{SE}$ is small and equation~\ref{eq:phi.ex1} provides a smooth approximation. 

As a numerical illustration, we consider four levels of true effect size: very high ($\theta = 2.5$), high ($\theta = 2.0$), medium ($\theta = 1.0$), and low ($\theta = 0.1$). We also consider four levels of between-experiment heterogeneity: low ($\sigma = 0.25$), medium ($\sigma = 0.50$), medium-high ($\sigma = 0.75$), and high ($\sigma = 1.50$). Table~\ref{tab:composition} reports $\mu$, and the implied $\rho$ for each pair of $(\theta,\sigma),$ together with the corresponding panel of Figure~1.

\begin{table}[ht]
\centering
\small
\renewcommand{\arraystretch}{1.25}
\begin{tabular}{lcccc}
\hline
& $\theta = 2.5$  & $\theta = 2.0$ 
& $\theta = 1.0$  & $\theta = 0.1$  \\
&(very high)&(high)&(medium)&(low) \\
\hline
\multicolumn{5}{l}{\textit{$\sigma = 0.25$ (low)}} \\
\hspace{.3cm}$\mu$ after heterogeneity & 0.992 & 0.974 & 0.834 & 0.539 \\
\hspace{.3cm}Implied $\rho$             & 0.004 & 0.009 & 0.027 & 0.037 \\
\hspace{.3cm}Figure~1 panel             & ${\sim}$A & ${\sim}$A & B & B \\
\hline
\multicolumn{5}{l}{\textit{$\sigma = 0.50$ (medium)}} \\
\hspace{.3cm}$\mu$ after heterogeneity & 0.987 & 0.963 & 0.814 & 0.536 \\
\hspace{.3cm}Implied $\rho$             & 0.028 & 0.049 & 0.102 & 0.128 \\
\hspace{.3cm}Figure~1 panel             & B & B & C & D \\
\hline
\multicolumn{5}{l}{\textit{$\sigma = 0.75$ (medium-high)}} \\
\hspace{.3cm}$\mu$ after heterogeneity & 0.977 & 0.945 & 0.788 & 0.532 \\
\hspace{.3cm}Implied $\rho$             & 0.091 & 0.130 & 0.204 & 0.234 \\
\hspace{.3cm}Figure~1 panel             & C & D & E & F \\
\hline
\multicolumn{5}{l}{\textit{$\sigma = 1.50$ (high)}} \\
\hspace{.3cm}$\mu$ after heterogeneity & 0.917 & 0.866 & 0.710 & 0.522 \\
\hspace{.3cm}Implied $\rho$             & 0.388 & 0.422 & 0.470 & 0.487 \\
\hspace{.3cm}Figure~1 panel             & Beyond F & Beyond F & Beyond F & Beyond F \\
\hline
\end{tabular}
\caption{$\mu$ and implied $\rho$ from population heterogeneity across labs under Normal error $\varepsilon_i \sim \mathrm {Nor}(0,\sigma^2)$.}
\label{tab:composition}
\end{table}

Table~\ref{tab:composition} is computed under $\mathrm{SE} = 1,$ or in other words, $\sigma_s = \sqrt{n}$ by convention. For large $n,$ $\mathrm{SE}$ is smaller, which leads to a test that is more sensitive to between-experiment differences in $\theta_i,$ which amplifies the effect of heterogeneity on $\phi_i$ across labs, increasing implied $\rho$. The values in Table~\ref{tab:composition} should therefore be read as conservative on the true $\rho$ implied by a given level of population heterogeneity $\sigma.$ For example, take $\theta = 1.0$ (medium effect) and $\sigma = 0.50$ (medium heterogeneity). The mean replicability rate is
$$\mu 
    = \Phi\!\left(\frac{\theta}{\sqrt{1 + \sigma^2}}\right)
    = \Phi(0.8944) \approx 0.814.
$$
We look at the quantitative effect of $n$ on $\rho$ in detail in Example~2, where finite sample size amplification is documented systematically.

For this same numerical example, the variance of $\phi$ is $\mathbb{E}[\phi_i^2] - \mu^2 \approx 0.0155,$ and we have 
$$\rho = \frac{\mathbb{E}[\phi_i^2] - \mu^2}{\mu(1-\mu)} \approx \frac{0.0155}{(0.814)(0.186)} \approx 0.102,$$
which maps to panel~C of Figure~1 (see Appendix for calculation of $\mathbb{E}[\phi_i^2]$).

Notable patterns emerge in Table~\ref{tab:composition}. Low heterogeneity ($\sigma = 0.25$) leaves effect sizes closer to the exact replication regime; we are in panel~A or B regardless of $\theta.$ When $\sigma = 0.50,$ weak and medium effects are pushed into panels C and D, where discriminability is already substantially compromised. At medium-high heterogeneity ($\sigma = 0.75$), the table spans panels C through F across the four effect sizes, covering the full problematic range of Figure~\ref{fig:variability.phi}. At high heterogeneity ($\sigma = 1.50$), every effect size lands beyond panel~F, where inference is not informative. 

We also notice that weaker effects reach higher $\rho$ panels under smaller heterogeneity than stronger effects do. At $\sigma = 0.50,$ a low effect size is already in panel~D while a very high effect size remains near panel~B. Strong effects enjoy a degree of natural discriminability that weak effects do not. Unfortunately, the results most actively disputed in the replication literature are precisely the weak to moderate ones, where $\rho$ climbs steeply even under these conservative assumptions about cross-replication variation. Demarcation based on replicability fails most severely for exactly the results it is most urgently invoked to adjudicate.

\subsection*{2. Effect of stimulus heterogeneity on replicability rate}
The second source of non-exactness we examine is delivery error---the gap between the stimulus an experimenter intends to administer and what they actually administer. This failure mode comes in two statistically distinct varieties. Under unsystematic error, which we call \textit{noise}, delivery varies randomly around the intended level. Under systematic error, which we call \textit{bias}\footnote{This is not the bias of an estimator.}, the experiment consistently over- or under-delivers the stimulus. These errors may arise due to protocol misreading, equipment miscalibration, or differences in how the stimulus is operationalized. Both are empirically well-documented and both matter for replicability, but in different ways and through different channels.

In psychophysics, stimulus calibration differences across labs are a recognized obstacle to replicability even under standardized protocols. \citet{ward2015achieving} demonstrate persistent cross-laboratory variation in psychophysical scaling across four labs, with failures of convergence arising even when the same protocol was followed. In pharmacology and preclinical biomedical research, the problem can be more severe. \citet{niepel2019multi} found up to 200-fold inter-center variability in drug dose-response estimates across five labs even after distributing identical reagents, cell lines, and experimental protocols. More broadly, \citet{begley2012raise} and~\citet{prinz2011believe} document replication failure rates of $89\%$ and $75\%,$ respectively, in landmark cancer biology studies, with dosage and reagent variation among the identified contributing factors. Our second example formalizes both sources of non-exactness within the psychometric function framework, making their separate and joint effects on replicability structurally explicit.

We assume that a sequence of experiments aims to administer a stimulus at intensity $s$ and record a binary response from each subject. The conditional probability, $\theta,$ of response taking value $1$ is modeled as a psychometric function $\theta = \Phi (u),$ of the standardized stimulus $u = (s-s_o)/\tau,$  where $s_o$ is the detection threshold and $1/\tau$, $(\tau \neq 0)$ is the slope controlling the steepness of the dose-response curve.

We consider non-exact replications when experiment $i$ delivers stimulus 
$$s_i = s + \delta_i,\;\;\delta_i \sim\mathrm {Nor}(b,\, \sigma^2),$$ 
where $\delta_i$ are independent and identically distributed with bias $b\neq 0.$ Experiments consistently over- or under-deliver the stimulus relative to the target value $s,$ and $\sigma^2$ is the variance in delivery. This is a measurement error model with nonzero mean in errors and it captures bias and noise. Thus, we have 
\begin{equation} \label{eq:theta}
 \theta_i = \Phi\left(u + \delta_i/\tau\right).
\end{equation}
For $m$ replications, the model can be seen as a generalized linear regression with binary outcome and probit link function $\Phi^{-1}$ at the level of experiments. 

We define the inferential problem  arbitrarily as testing the hypothesis $H_0: \theta\leq 0.5$ and the result of interest as rejection of $H_0$ indicated by $\mathbf{1}_{\{\hat{\theta} > c\}},$ where $\hat{\theta}$ is the random proportion of subjects whose response is $1,$ and $c$ is fixed and known. Under non-exact replications with large sample size $n,$ the true replicability rate for experiment $i$ is $$\phi_i = P(\hat{\theta}_i\geq c) = \Phi\left(\frac{\theta_i - c}{\sqrt{\theta_i(1-\theta_i)/n}}\right),$$ by Normal approximation to Binomial distribution. As the sample size $n \rightarrow \infty$ in each experiment, by equation~\ref{eq:theta}, the distribution of $\hat{\theta}_i$ converges to $\Phi(u + \delta_i/\tau)$ and the replicability rate of each experiment $\phi_i$ converges to $0$ if $\Phi(u + \delta_i/\tau)<c$ and to $1$ otherwise. In this limit, all the randomness in replicability rates across experiments comes purely from between-experiment delivery variation, and within-experiment sampling plays no role. For large but finite $n,$ we use approximation $\phi_i = \Phi(u + \delta_i/\tau)$ in Table~\ref{tab:example2_extended}, which treats the continuous response probability as the replicability rate, where within-experiment sampling variability is negligible.

The mean replicability rate $\mu$ of the Betabinomial is 
\begin{equation*}
    \mu = \Phi\!\left(\frac{u+b/\tau}{\sqrt{1 + (\sigma/\tau)^2}}\right),
\end{equation*}
where ratio $\sigma/\tau$ is the scaled non-exactness parameter. Intuitively, a shallow dose-response curve characterized by large $\tau$ provides natural robustness to delivery noise, while a steep curve characterized by small $\tau$ amplifies it. The intraclass correlation parameter is given by $\rho = \mathbb{V}(\phi_i)/\mu(1-\mu).$

As a numerical illustration, we fix $u = 1$, corresponding to a response probability of $\Phi(1) = 0.841$ under exact replication. We consider seven levels of scaled bias: 
$$b/\tau \in \{-1.5, -1.0, -0.5, 0.0, 0.5, 1.0, 1.5\},$$
spanning under- to over-delivery. We fix scaled delivery noise at four levels: 
$$\sigma/\tau \in \{0.25, 0.50, 0.75, 1.50\}.$$
Table~\ref{tab:example2_extended} reports $\mu$ and the implied $\rho$ for each $(b/\tau,\sigma/\tau)$ under the large-$n$ approximation as well as $n=100$ and the corresponding panel in Figure~\ref{fig:variability.phi}. For $n = 100,$ we use the power of one-sided exact binomial test $H_0:\theta \leq 0.5$ which rejects at $c=0.59,$ at $\alpha = 0.044$, making the dependence of $\phi_i$ on the experiment's sample size explicit. The two rows within each bias block show how the choice of approximation affects the implied severity of non-exactness.

\begin{table}[htbp]
\centering
\footnotesize
\setlength{\tabcolsep}{4pt}
\begin{tabular}{rr rr r  rr r  rr r  rr r}
\toprule
& & \multicolumn{3}{c}{$\sigma/\tau = 0.25$}
  & \multicolumn{3}{c}{$\sigma/\tau = 0.50$}
  & \multicolumn{3}{c}{$\sigma/\tau = 0.75$}
  & \multicolumn{3}{c}{$\sigma/\tau = 1.50$} \\
\cmidrule(lr){3-5}\cmidrule(lr){6-8}\cmidrule(lr){9-11}\cmidrule(lr){12-14}
$b/\tau$ & $n$ & $\mu$ & $\rho$ & Panel
                & $\mu$ & $\rho$ & Panel
                & $\mu$ & $\rho$ & Panel
                & $\mu$ & $\rho$ & Panel \\
\midrule

\multirow{2}{*}{$+1.5$}
 & $\infty$ & 0.992 & 0.004 & ${\sim}$A & 0.987 & 0.028 & B      & 0.977 & 0.091 & C      & 0.917 & 0.388 & ${>}$F \\
 & 100      & 1.000 & ---   & ---       & 1.000 & 0.414 & ${>}$F & 0.999 & 0.697 & ${>}$F & 0.936 & 0.901 & ${>}$F \\
\addlinespace

\multirow{2}{*}{$+1.0$}
 & $\infty$ & 0.974 & 0.009 & ${\sim}$A & 0.963 & 0.049 & B      & 0.945 & 0.130 & D      & 0.866 & 0.422 & ${>}$F \\
 & 100      & 1.000 & ---   & ---       & 1.000 & 0.513 & ${>}$F & 0.991 & 0.749 & ${>}$F & 0.882 & 0.910 & ${>}$F \\
\addlinespace

\multirow{2}{*}{$+0.5$}
 & $\infty$ & 0.927 & 0.018 & ${\sim}$A & 0.910 & 0.076 & C      & 0.885 & 0.170 & D      & 0.797 & 0.449 & ${>}$F \\
 & 100      & 1.000 & 0.106 & C         & 0.994 & 0.618 & ${>}$F & 0.955 & 0.795 & ${>}$F & 0.803 & 0.917 & ${>}$F \\
\addlinespace

\multirow{2}{*}{$0.0$}
 & $\infty$ & 0.834 & 0.027 & B         & 0.814 & 0.102 & C      & 0.788 & 0.204 & E      & 0.710 & 0.470 & ${>}$F \\
 & 100      & 0.997 & 0.296 & ${>}$F    & 0.936 & 0.713 & ${>}$F & 0.849 & 0.829 & ${>}$F & 0.699 & 0.922 & ${>}$F \\
\addlinespace

\multirow{2}{*}{$-0.5$}
 & $\infty$ & 0.686 & 0.035 & B         & 0.673 & 0.121 & C      & 0.655 & 0.226 & F      & 0.609 & 0.483 & ${>}$F \\
 & 100      & 0.846 & 0.537 & ${>}$F    & 0.710 & 0.770 & ${>}$F & 0.646 & 0.848 & ${>}$F & 0.575 & 0.924 & ${>}$F \\
\addlinespace

\multirow{2}{*}{$-1.0$}
 & $\infty$ & 0.500 & 0.037 & B         & 0.500 & 0.128 & D      & 0.500 & 0.234 & F      & 0.500 & 0.487 & ${>}$F \\
 & 100      & 0.222 & 0.560 & ${>}$F    & 0.339 & 0.774 & ${>}$F & 0.389 & 0.849 & ${>}$F & 0.443 & 0.924 & ${>}$F \\
\addlinespace

\multirow{2}{*}{$-1.5$}
 & $\infty$ & 0.314 & 0.035 & B         & 0.327 & 0.121 & C      & 0.345 & 0.226 & F      & 0.391 & 0.483 & ${>}$F \\
 & 100      & 0.005 & 0.335 & ${>}$F    & 0.083 & 0.724 & ${>}$F & 0.174 & 0.833 & ${>}$F & 0.318 & 0.922 & ${>}$F \\

\bottomrule
\end{tabular}
\caption{Effect of stimulus delivery bias ($b/\tau$) and noise ($\sigma/\tau$) on mean replicability rate $\mu$ and intraclass correlation $\rho,$ for two specifications of $n$. Fixed standardized stimulus $u = (s-s_o)/\tau = 1.0$ which corresponds to replicability rate $\Phi(1) = 0.841$. Row $n \rightarrow\infty$: large-$n$ approximation $\phi_i = \Phi(u + \delta_i/\tau)$. Row $n = 100$: exact power of one-sided binomial test $H_0:\theta \le 0.5$, with the result defined as $1$ if the test rejects the null hypothesis which corresponds to $c= 0.59$ at $\alpha = 0.044$. Panel labels follow Figure~1. Entries marked \textup{---} correspond to $\mu = 1$, making $\rho = \mathbb{V}(\phi_i)/[\mu(1-\mu)]$ undefined.}
\label{tab:example2_extended}
\end{table}
 
Across all cells, the large-$n$ approximation understates $\rho$ throughout, for the test specification used here. This is expected since within-experiment sampling variability is negligible. Across all cells, the finite-$n$ $\rho$ is substantially larger than the corresponding large-$n$ value, often by an order of magnitude. At $b/\tau = 0$ and $\sigma/\tau = 0.50$, the large-$n$ approximation gives $\rho = 0.102$ (panel~C), while $n = 100$ gives $\rho = 0.713$ (beyond panel~F). The reason is that the significance test converts continuous between-lab variation in true response probabilities into binary outcomes conditional on $n,$ amplifying the variance of $\phi_i$ across experiments. The large-$n$ approximation does not take into account this source of variability.

When bias is positive and experiments deliver a stimulus well above the detection threshold, the test passes with near certainty at $n = 100$, giving $\mu \approx 1$ in the finite-$n$ row. When bias is strongly negative, experiments produce results consistently below the critical value and $\mu$ shrinks. The most extreme case considered is $b/\tau = -1.5$, $\sigma/\tau = 0.25$, where the large-$n$ row gives $\mu = 0.314$, while the finite-$n$ row gives $\mu = 0.005$. The large-$n$ approximation does not just understate $\rho,$ it also conceals how severely systematic under-delivery degrades the apparent replicability rate.
 
In our example, positive and negative bias of the same magnitude do not produce symmetric $\rho$ values. Consistent under-delivery yields higher $\rho$ than consistent over-delivery. Since $\phi_i = \Phi(u + \delta_i/\tau),$ the sensitivity of $\phi_i$ to delivery variation is governed by the standard Normal density evaluated at the bias-shifted stimulus $u + b/\tau.$ The density is largest near $0$ and decays away from it. Over-delivery increases $u + b/\tau$ above $u = 1.0,$ moving into a region where the standard Normal density is small. Delivery noise $\sigma/\tau$ produces little variation in $\phi_i$ across experiments, keeping $\rho$ low. Under-delivery, on the other hand, decreases $u + b/\tau$ toward zero, where the density is larger. The same noise now produces much greater variation in $\phi_i,$ inflating $\rho.$ 

The two sources of non-exactness, bias and noise, do not operate independently once within-experiment sampling variability is taken into account. In the $n \rightarrow\infty$ rows, $\mu$ is primarily controlled by $b/\tau$ and $\rho$ is primarily controlled by $\sigma/\tau$, so the two sources of non-exactness are at least partially separable. There is no such observable separation at $n = 100$. Finite-$n$ $\rho$ is large across virtually all cells regardless of bias level, because the significance test amplifies between-experiment variation into binary outcomes for any $n$ large enough to resolve differences among experiments. The finite-$n$ $\mu$ responds strongly to both bias and noise together, since power depends on the distribution of $\hat{\theta}_i$ relative to the critical value. Positive bias can therefore inflate $\mu$ and give an appearance of strong replicability, while simultaneously producing a high $\rho$ and making it difficult to discriminate between genuinely high- and low-replicability sequences. 

Therefore, bias shifts $\mu$ away from the value that exact replication would produce, making the estimated replicability rate an unreliable summary of the underlying effect, and might give us unearned confidence in the replication results. This creates the risk of getting trapped in what~\citet[][p.6]{meng2020reproducibility} calls the \textit{replication paradox}: ``the more we replicate, the surer we fool ourselves''. On the other hand, noise inflates $\rho$ and degrades discriminability regardless of where $\mu$ sits. Total delivery heterogeneity distorts replicability rate and undermines the capacity of any replication sequence to distinguish results that hold from results that do not.
\section*{Estimating the replicability rate of a result: Many~Labs~4}
The numerical examples above show how discriminability is lost when $\rho$ can be traced to a well-defined physical or statistical source. Real experiments often have untraceable sources of variability. The heterogeneity introduced across a replication sequence is typically a compound of procedural, sampling, measurement, and contextual differences that in practice resist reduction to a single parameter and may not be jointly quantified. What the formal framework characterizes as $\rho > 0$ in practice, is the accumulated weight of many decisions that differ across experiments. In this section, we examine one such observed sequence in detail, to give a sense of what that weight actually looks like and why the discriminability problem is understated in theory.

Our case is Many~Labs~4~\citep{klein2022many} (ML4), a meta-study involving multiple replications of Study~1 of~\citet{greenberg1994role}, which reported a foundational Terror Management Theory result. The treatment in the reference study included a mortality salience manipulation by prompting participants to write about their own death. Mortality salience was found to result in a heightened worldview defense, which was operationalized as increased preference for a pro-American over an anti-American essay author. We chose the~\citet{klein2022many} study both because it includes a clear non-exact replication sequence around a single result and because it explicitly manipulates design heterogeneity in replications, allowing for an opportunity to further contextualize $\rho$. Labs were randomized to either design their replication independently from the published method section (In House, IH), or receive direct procedural guidance from the original authors (Author Advised, AA). The AA protocol was meant to reduce design heterogeneity. The reference result was successfully replicated in 1 of 17 labs and positive effects were obtained in 7 of them.

\subsection*{Qualitative evaluation} The procedural differences between the reference study and its replications are extensive and cut across virtually every component of the experimental design. They also vary substantially among the replications themselves, as documented in Table~2 of~\citet{klein2022many}. What follows is a catalog by category. It is not meant to be exhaustive, but sufficient for illustration purposes.
 
\textit{Treatment.} The reference study used five between-subject conditions: subtle own-death salient, subtle other's-death salient, deeper own-death salient, deeper other's-death salient, and television salient. ML4 retained only the first (as treatment) and last (as control) of these.

\textit{Sample size and composition.} The reference recruited 58 introductory psychology students at a single American university, tested in groups of three to five, with 11 to 12 participants per condition and the experiment was performed in 1994. 

ML4 enrolled 1,578 participants after exclusions across 17 geographically and institutionally diverse labs, with sample sizes varying between 58 and 141 participants across sites. The replication samples spanned different institutions, regions, and in some cases demographic profiles, and were performed in 2016-2017. These are not replications of the same experiment. 

\textit{Procedure, materials, and timing.} When presented to the subjects, the reference study was ostensibly introduced as two short studies (to prevent subjects from explicitly connecting the manipulation to the dependent measure), which they received in individual cubicles to ensure privacy. The so-called ``first study'' began with two filler measures, which are not specified in~\citep{greenberg1994role}, but were presumably included to obscure the purpose of the mortality salience manipulation that followed. Subjects were assigned to one of four treatment and one control conditions. Each condition included two open-ended questions where subjects were asked to describe their emotions and imagine what would happen if the event included in their respective condition were to happen. For example, for the subtle own-death salience treatment, subjects were asked the following two questions: ``Please briefly describe the emotions that the thought of your own death arouses in you” and ``Jot down, as specifically as you can, what you think will happen to you physically as you die and once you are physically dead.’’ Wording was adjusted accordingly per condition. Subjects next responded to two open-ended statements, completed a 10-item mood assessment, and the 20-item PANAS mood scale\footnote{The Positive and Negative Affect Schedule is a widely used self-report questionnaire of positive and negative dimensions of mood. Respondents rate a series of mood-related adjectives on a 5-point Likert scale, indicating how much they feel that way within a specified time frame~\citep{watson1988development}.}. After this first section was concluded, subjects received a cover story to introduce the ``second study’’, which included the dependent measure. Basically, they were asked to read two brief essays written by foreign students, one pro-American and one anti-American, and answered three questions to evaluate the author of the essay and two questions to evaluate the essay itself. Finally, the subjects were debriefed.

ML4 was also presented as two separate studies. The AA protocols were administered in person in cubicles, using paper and pencil, as in the reference study while IH protocols were administered on the computer, either in the lab or online, depending on the site. In the ``first study,’’ at all but three sites, subjects also started with filler measures. At AA sites, these included three sets of measures. At IH sites, the filler measures differed in number and content. The manipulation followed the filler measures. Subjects were randomly assigned to one of two conditions, treatment or control. The manipulations of these conditions included the same corresponding open-ended questions as in the reference study. More filler measures followed. Subjects next responded to PANAS in 7 of the 10 IH labs and PANAS-X\footnote{PANAS-X is an extended version of the PANAS, with a specific subscale allowing researchers to distinguish among negative and positive emotions.} in all AA labs. AA protocol next introduced two extended measures as additional filler tasks to provide more time between the treatment and the dependent measure, and the first part was concluded. At 4 of the IH sites, no filler task was provided. At other IH sites, different filler tasks were presented than those in the AA protocol. Next, a so-called ``unrelated second study’’ was presented, which included two essays as in the reference study. However, in the AA protocol, the anti-American essay was revised to be more forceful and extreme. Emerging stimulus heterogeneity across labs resembles the situation described in our second numerical example. Half of the IH sites used the unchanged essays from the reference study, while the other half used different essays. As a dependent measure, subjects were asked to rate the same five evaluation questions as in the reference study. 

\textit{Experimenter behavior and physical environment.} The AA protocol included explicit instructions not described in the reference such as selecting relaxed research assistants, using a covered box for packet submission to ensure a feeling of confidentiality, and having experimenters dress and act casually. The reference experiment's method section contains none of these specifications, meaning the IH condition could not have matched them, and there is no way to know whether the reference study's testing environment implicitly embodied any of them.

\textit{Exclusion criteria.} The reference applied no post-hoc exclusions, keeping all subjects in the data analysis. ML4 applied multiple exclusion criteria including minimum cell size thresholds, completion checks on both writing prompts and all essay ratings, and in the AA condition, nationality-based exclusions. These choices structure who is counted in the analysis in ways that have no direct counterpart in the reference. The variation in sampled populations, along with the varying exclusion criteria, presents a close real life counterpart to our first numerical example.

\textit{Analyses.} In the reference, a one-way analysis of variance comparing four treatments and one control was performed on the composite measures of preference for the pro-American essay author. In the replications, an independent samples t-test comparing one treatment and one control was performed for each lab.  

Joint effect of these differences is challenging to to $\rho$ quantitatively. In the vocabulary of the benchmark model, each site realized a different instantiation of a theory that specifies almost none of these conditions. This is precisely the situation our framework characterizes as intractable for replicability-based demarcation. When the sources of non-exactness are qualitatively diverse, compound, and unmeasurable, the mean replicability rate $\mu$ loses its interpretive foundation as a diagnostic quantity, and the binary verdict of ``replicated'' or ``not replicated'' cannot be read as a stable property of the underlying result. It can only be read as a property of this particular sequence, under these particular conditions, at this particular moment.

\subsection*{Quantitative evaluation} 

The qualitative catalog above establishes that the ML4 sequence involves non-exactness across virtually every dimension of experimental design. What it cannot do is assign a number to that non-exactness. We now attempt a partial quantitative characterization, but the attempt itself must be situated carefully with respect to the two models developed in the statistical theory section.

Under the operational model at all $k = 1,$ the binary replication verdicts carry no information about $\rho;$ the marginal likelihood factors as a product of $\mathrm{Bernoulli}(\mu)$ terms and $\rho$ drops out entirely. No meaningful inference about the degree of non-exactness is possible from the binary verdict data alone. To make any quantitative progress, we must therefore step outside the operational model and use a different data structure. So we will use the continuous Hedges' $g$ values reported at each ML4 site\footnote{The result type $R$ varies across the numerical examples and the ML4 analysis. Example~1 defines replication as a directional result $\theta_i > 0$ assuming $SE=1$; Example~2 as rejection of a one-sided binomial test; the ML4 analysis uses the finite-$n$ correction. These are not interchangeable targets, and the specific values of $\mu$ and $\rho$ depend on the choice of $R,$ and the within-experiment sample size. What is invariant across all choices of $R$ is the theoretical machinery. The variance formula, the irreducible floor, and the non-identifiability of $\rho$ at $k = 1$ hold for any fixed binary result type. The examples use specific but representative choices of $R$ to show that realistic experimental conditions map to $(\mu, \rho)$ where discriminability failure and false precision already apply.}.

We use an auxiliary model (a Normal random-effects model) that aims to obtain a $\rho$ value that can then be plugged into the benchmark model to illustrate what discriminability looks like at the inferred level of non-exactness. We follow a three-step chain. First, we fit a $\mathrm {Nor}(\theta, \sigma^2)$ model to the 18 Hedges' $g$ values and obtain the posterior for $(\theta, \sigma).$ Second, we transform that posterior to the posterior of $(\mu, \rho)$ via $\phi_i = \Phi(\theta_i/(SE_i)),$ with finite sample correction using site-specific standard errors $SE_i$. Third, we locate the posterior interval bounds of $(\rho)$ on Figure~\ref{fig:variability.phi}.

This procedure ignores some uncertainty. The Hedges' $g$ values are themselves noisy estimates of the true $\theta_i.$ We do not pursue a full meta-analytic approach here because the auxiliary model serves a diagnostic rather than inferential purpose. It locates the ML4 sequence on the non-exactness landscape of Figure~\ref{fig:variability.phi} rather than providing a definitive estimate of between-experiment heterogeneity. Separately, effect size heterogeneity does not capture all sources of design heterogeneity. Some determinants of $\rho,$ such as variation in sample composition, measurement error, and ceiling effects, affect $\phi_i$ without necessarily changing the estimated effect size, and are not reflected in the $g_i$ values at all. 

Table~3 of~\citet{klein2022many} reports Hedges' $g$ for each of 17 replication sites (under the first set of exclusion criteria). The reference study~\citet{greenberg1994role} reported Cohen's $d = 1.34$ for the pairwise comparison between the subtle own-death ($n_1 = 12$) and television-control conditions ($n_2 = 11$). To place the reference on the same scale as the replications, we convert to Hedges' $g$ using the small-sample correction factor $ J = 1 - 3/(4\nu - 1),$ where $\nu = n_1 + n_2 - 2$ is the degrees of freedom of the effect size estimate. For the reference experiment, $\nu = 12 + 11 - 2 = 21,$ giving $J = 1 - 3/83 \approx 0.964,$ and hence $g = Jd = (0.964)(1.34) \approx 1.29.$ The correction deflates $d$ to remove the positive small-sample bias inherent in Cohen's $d$ as an estimator of the population standardized mean difference~\citep{hedges1981distribution}. All 18 values (one reference, 17 replications) are now on a common scale and can be treated as draws from the same distribution (see Table~\ref{tab:Hedges}). 
\begin{table}[htbp]
\begin{center}
\begin{tabular}{lcc}
\toprule
Quantity & ML4 only & Reference + ML4 \\
\midrule
Mean $g$ & $0.055$ & $0.123$ \\
SD $g$   & $0.250$ & $0.379$ \\
$m$             & $17$    & $18$    \\
\bottomrule
\end{tabular}
\caption{Means and standard deviations of estimated effect sizes from ML4 replication studies, with and without the reference.}
\label{tab:Hedges}
\end{center}
\end{table}

\noindent
\textbf{Posterior inference.} In the auxiliary effect size model, each experiment contributes one true effect size $\theta_i$ drawn from $\mathrm {Nor}(\theta, \sigma^2).$ 

To obtain the implied posterior distribution of the mean replicability rate $\mu$ and the heterogeneity parameter $\rho$, we propagate posterior uncertainty in $(\theta,\sigma^2)$ by Monte Carlo simulation. For each $s \in \{1,\cdots,300000\},$ we draw $(\theta^{(s)},\sigma^{2(s)})$ and simulate site-level effects $\theta_i^{(s)} \sim \mathrm {Nor}(\theta^{(s)},\sigma^{2(s)})$ for $i = 1,\dots,m.$ We use the transformation $\phi_i^{(s)} = \Phi(\theta_i^{(s)}/\mathrm{SE}_i)$ and compute
\begin{equation}\label{eq:mu.tranform}\mu^{(s)} = \frac{1}{m} \sum_{i=1}^m \phi_i^{(s)}, \qquad
\rho^{(s)} = \frac{\mathbb{V}(\phi_i^{(s)})}
                  {\mu^{(s)}\bigl(1-\mu^{(s)}\bigr)}.
\end{equation} Posterior summaries of $\mu$ and $\rho$ are obtained from the distribution of $(\mu^{(s)},\rho^{(s)}).$ The priors are $\theta \mid \sigma^2 \sim \mathrm {Nor}(\mu_0, \sigma^2/\kappa_0)$ and $\sigma^2 \sim \mathrm {Inv\text{-}Gamma}(\alpha_0, \beta_0).$ We use two hyperparameter specifications. The first is Jeffreys prior, which is improper, with Uniform density on $\theta$ and $\log \sigma,$ and therefore carries minimal information about where the mean effect sits. The second set of priors serves as a sensitivity check. For this purpose, we use a weakly informative prior that places mild prior mass near zero effect size (see the Appendix for details).

Tables~\ref{tab:ML4} and~\ref{tab:AAIH} report posterior means and $95\%$ HDIs for $\mu$ and $\rho.$ The credible intervals of $\mu$ quantify posterior uncertainty about the mean replicability rate, given the estimated effect sizes under the auxiliary Normal model. The credible intervals of $\rho$ quantify posterior uncertainty about the degree of non-exactness implied by the dispersion of those effect sizes. The panel range column translates the $\rho$ interval into the corresponding panels of Figure~\ref{fig:variability.phi}, showing which non-exactness regimes are consistent with the data.


\begin{table}[htbp]
\begin{center}
\small
\setlength{\tabcolsep}{5pt}
\begin{tabular}{lc l rr r}
\toprule
Case & $m$ & Prior
  & $\mu$ (95\% HDI)
  & $\rho$ (95\% HDI)
  & Panel range \\
\midrule

\multirow{2}{*}{ML4 only}
  & \multirow{2}{*}{17}
  & Jeffreys
  & $0.565\ [0.364,\, 0.765]$
  & $0.373\ [0.175,\, 0.576]$
  & D -- ${>}$F \\[2pt]
  & & Weakly inf.
  & $0.543\ [0.307,\, 0.777]$
  & $0.551\ [0.335,\, 0.764]$
  & ${>}$F \\
\addlinespace

\multirow{2}{*}{Reference $+$ ML4}
  & \multirow{2}{*}{18}
  & Jeffreys
  & $0.608\ [0.384,\, 0.820]$
  & $0.505\ [0.291,\, 0.725]$
  & ${>}$F \\[2pt]
  & & Weakly inf.
  & $0.585\ [0.351,\, 0.815]$
  & $0.591\ [0.377,\, 0.796]$
  & ${>}$F \\

\bottomrule
\end{tabular}
\caption{Posterior means and 95\% credible intervals of $\mu$ and $\rho$ from observed heterogeneity in Hedges' $g$ across ML4 studies, with and without the reference study, using the finite-$n$ mapping. The model is $g_i \sim \mathrm {Nor}(\theta, \sigma^2)$ independently. Two prior specifications are compared under the Normal-Inverse-Gamma model: Jeffreys (improper, $\kappa_0 \to 0$, $\alpha_0 \to 0$, $\beta_0 \to 0$) and weakly informative ($\mu_0 = 0$, $\kappa_0 = 1$, $\alpha_0 = 1$, $\beta_0 = 1$). Posterior summaries are obtained using 300{,}000 draws from the Normal-Inverse-Gamma posterior by direct sampling. $\mu$ and $\rho$ are derived from each posterior draw via the Monte Carlo simulation given by equation~\ref{eq:mu.tranform}. The panel range column maps the 95\% credible interval for $\rho$ to the panels of Figure~\ref{fig:variability.phi}.}
\label{tab:ML4}
\end{center}
\end{table}

In Table~\ref{tab:ML4}, posterior means of $\mu$ range from $0.543$ to $0.608$ and the two priors produce intervals that largely overlap. The estimates of $\rho$ are more interesting. For ML4 only ($m = 17$), the posterior mean of $\rho$ is $0.373$ under Jeffreys and $0.551$ under the weakly informative prior, placing the sequence in panels D--F and beyond panel F, respectively. For Reference $+$ ML4 ($m = 18$), $\rho$ is $0.505$ under Jeffreys and $0.591$ under the weakly informative prior, both beyond panel F. The weakly informative prior consistently yields higher $\rho$ than Jeffreys, reflecting the prior's leverage on $\sigma$ when $m$ is small.

The reference experiment's $g = 1.29$ is an outlier relative to the replication sequence and substantially influences the joint posterior. Under Jeffreys prior, including the reference raises the posterior mean of $\rho$ from $0.373$ to $0.505.$ Under the weakly informative prior the corresponding shift is from $0.551$ to $0.591.$ In all cases the $\rho$ intervals span panels D through beyond F, placing the sequence in a severe non-exactness regime regardless of whether the reference study is included.

It is instructive to consider what the posteriors would look like if non-exactness were assumed away. If a researcher fixes $\rho = 0$ and uses the Binomial likelihood in place of the Betabinomial, the posterior of $\mu$ would concentrate tightly around the estimated replication proportion, because all variability is attributed to sampling noise and none to between-experiment heterogeneity. However, assuming $\rho = 0$ does not eliminate true heterogeneity. It merely hides it, reproducing the false sense of precision identified earlier.
\begin{table}[htbp]
\begin{center}
\small
\setlength{\tabcolsep}{5pt}
\begin{tabular}{lc l rr r}
\toprule
Group & $m$ & Prior
  & $\mu$ (95\% HDI)
  & $\rho$ (95\% HDI)
  & Panel range \\
\midrule

\multirow{2}{*}{AA only}
  & \multirow{2}{*}{7}
  & Jeffreys
  & $0.616\ [0.345,\, 0.868]$
  & $0.253\ [0.026,\, 0.534]$
  & B -- ${>}$F \\[2pt]
  & & Weakly inf.
  & $0.550\ [0.163,\, 0.903]$
  & $0.605\ [0.238,\, 1.000]$
  & F -- ${>}$F \\
\addlinespace

\multirow{2}{*}{IH only}
  & \multirow{2}{*}{10}
  & Jeffreys
  & $0.537\ [0.262,\, 0.808]$
  & $0.409\ [0.138,\, 0.687]$
  & D -- ${>}$F \\[2pt]
  & & Weakly inf.
  & $0.522\ [0.209,\, 0.830]$
  & $0.591\ [0.297,\, 0.878]$
  & ${>}$F \\
\addlinespace

\multirow{2}{*}{AA $+$ reference}
  & \multirow{2}{*}{8}
  & Jeffreys
  & $0.668\ [0.350,\, 0.973]$
  & $0.513\ [0.106,\, 0.863]$
  & C -- ${>}$F \\[2pt]
  & & Weakly inf.
  & $0.618\ [0.269,\, 0.952]$
  & $0.620\ [0.228,\, 0.976]$
  & F -- ${>}$F \\
\addlinespace

\multirow{2}{*}{IH $+$ reference}
  & \multirow{2}{*}{11}
  & Jeffreys
  & $0.608\ [0.319,\, 0.889]$
  & $0.542\ [0.241,\, 0.828]$
  & F -- ${>}$F \\[2pt]
  & & Weakly inf.
  & $0.582\ [0.279,\, 0.875]$
  & $0.619\ [0.336,\, 0.894]$
  & ${>}$F \\

\bottomrule
\end{tabular}
\caption{Posterior means and 95\% credible intervals of $\mu$ and $\rho$ by AA and IH protocol groups, with and without the reference study, using the finite-$n$ mapping. Model and priors are as in Table~\ref{tab:ML4}. Under the weakly informative prior, small-$m$ cases show wider $\rho$ intervals reflecting prior leverage. The wide $\rho$ intervals throughout reflect the amplification of between-site heterogeneity under finite sample sizes.}
\label{tab:AAIH}
\end{center}
\end{table}


We can use the AA/IH design manipulation to check whether the standardization effort reduced $\rho$ as intended (Table~\ref{tab:AAIH}). Under Jeffreys prior, the AA sequence has a posterior mean  $\rho = 0.253$ with $95\%$ credible interval $[0.026, 0.534],$ spanning panels B through beyond F. The IH sequence has posterior mean $\rho = 0.409$ with interval $[0.138, 0.687],$ spanning panels D through beyond F. The credible intervals overlap substantially, and both sequences are in severe non-exactness regimes. To assess whether the AA protocol meaningfully reduced $\rho,$ we compute the posterior distribution of the difference $\rho_{\mathrm{IH}} - \rho_{\mathrm{AA}}$ directly. The posterior mean of the difference is $0.156$ and the $95\%$ HDI is $[-0.246, 0.550],$ which includes zero. The probability that $\rho_{\mathrm{IH}} > \rho_{\mathrm{AA}}$ is $0.789,$ indicating a tendency toward higher heterogeneity in the IH sequence but with substantial uncertainty. Whatever reduction in design heterogeneity the AA protocol achieved, it was not sufficient to move the sequence out of the high-$\rho$ regime, making discriminability essentially impossible.


For the AA sequence, the posterior uncertainty about $\mu$ is substantial. The $95\%$ credible interval is $[0.345, 0.868]$ under Jeffreys prior, and $[0.163, 0.903]$ under the weakly informative prior. For the IH sequence, the corresponding intervals are $[0.262, 0.808]$ and $[0.209, 0.830].$ These intervals are consistent with mean replicability rates ranging from below $0.20$ to above $0.85,$ essentially covering the full range of scientifically meaningful values. Inference about $\mu$ from a single realization of either sequence provides very weak information about where the true mean replicability rate lies.
\vspace{0.2cm}

\begin{figure}[h] 
\begin{center}
\includegraphics[width=\textwidth]{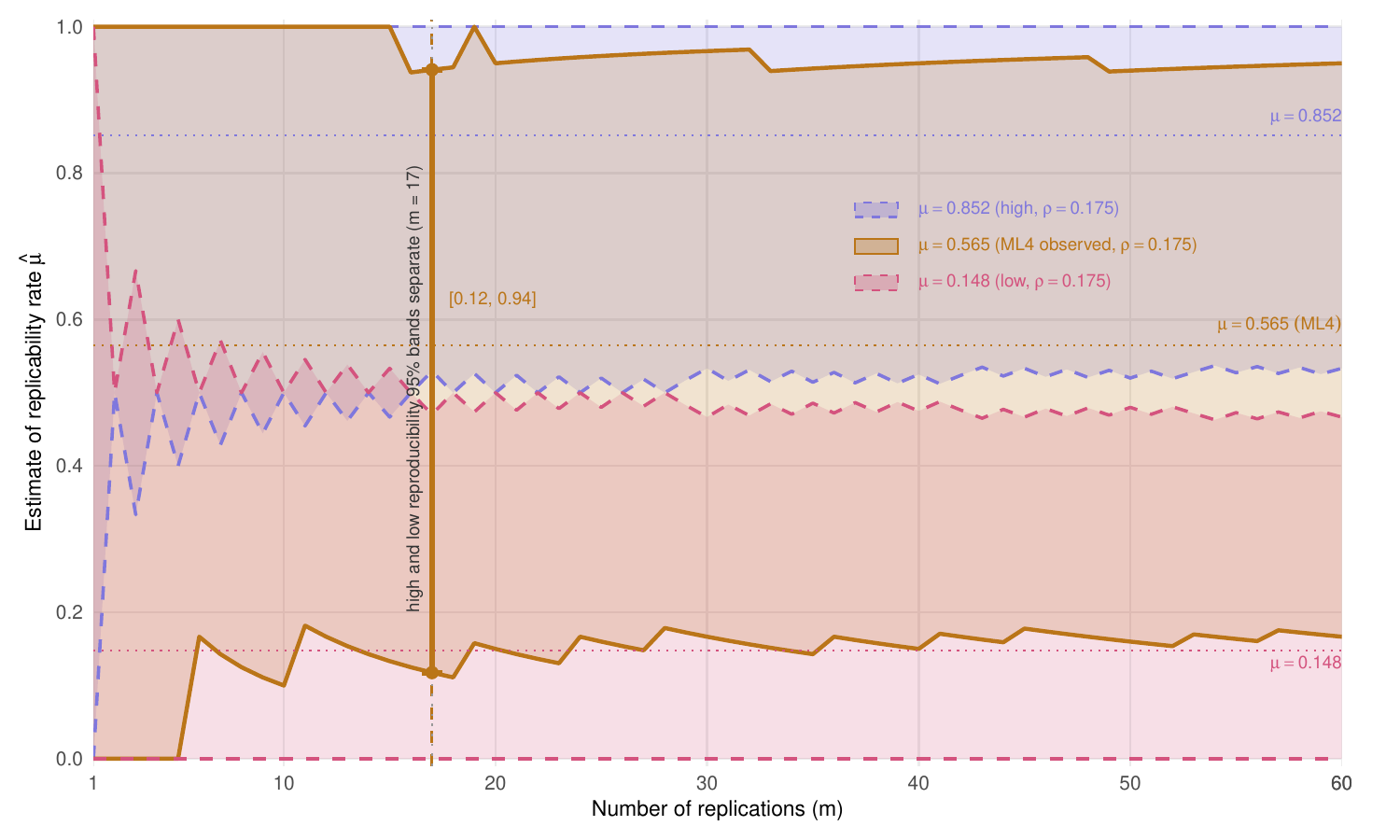}
\end{center}
\caption{Sampling distribution of $\hat{\mu}$ under the benchmark
model for the ML4 replication sequence ($m = 17$ sites, ML4 only). $95\%$ HDI of Betabinomial distribution as a function of $m,$ computed at $\rho = 0.175,$ the lower bound of the $95\%$ credible interval under Jeffreys prior (Table~\ref{tab:ML4}). Purple: $\mu = 0.852.$ Amber: ML4 posterior mean ($\mu = 0.565$). Pink: $\mu = 0.148.$ Purple and pink values are the most proximate symmetric pair that achieves separation at the actual ML4 sample size under this $\rho.$ Vertical amber line marks $m = 17.$ The jagged band boundaries reflect the discrete support of $\hat{\mu} = X/m$ under the binary verdict data.}
\label{fig:ml4}
\end{figure}

\noindent
\textbf{Theoretical expectation. }
Figure~\ref{fig:ml4} displays the $95\%$ HDI of $\hat{\mu}$ as a function of the number of replications $m,$ under the benchmark model using the ML4 only sequence. The figure is constructed under $\rho = 0.175$ (Table~\ref{tab:ML4}), which is the lower bound of the $95\%$ credible interval obtained under Jeffreys prior. We chose to plot the lower bound to show the highest discriminability scenario consistent with the data. Even this lower bound is substantially above panel D of Figure~\ref{fig:variability.phi}.

Three bands are shown. The amber band uses the ML4 posterior mean $\mu = 0.565.$ The purple and pink bands indicate what the most extreme $\mu$ values are that would just barely separate from each other at $m = 17.$ The answer is $\mu = 0.852$ (purple) and $\mu = 0.148$ (pink)---a pair whose $95\%$ HDI bands are $[0.529, 1.000]$ and $[0.000, 0.471],$ separated by a gap of $0.059.$ These values replace the earlier $(0.20, 0.80)$ pair used in the exact replication baseline of Figure~\ref{fig:variability.phi}, which does not separate at $m = 17$ under $\rho = 0.175$ or even asymptotically. The choice of $(0.148, 0.852)$ is not arbitrary. It is the most proximate symmetric pair that achieves separation at the actual ML4 sample size under this $\rho.$ The amber band at $m = 17$ is the 95\% incredible interval which spans $[0.12, 0.94],$ covering nearly the entire support of $\mu$. It contains both reference values $\mu = 0.148$ and $\mu = 0.852,$ consistent with a result that almost certainly does not replicate and equally consistent with one that almost certainly does. No inference about the true replicability rate of the mortality salience effect is possible from this sequence alone.

The effective sample size at $m = 17$ is $m_e \approx 4.5$ under $\rho = 0.175$ and $m_e \approx 2.4$ under the posterior mean $\rho = 0.373.$ Rather than careless design or insufficient effort, the wide amber band is a result of $\rho > 0,$ the non-exactness inherent in any multi-site replication sequence that cannot hold all experimental conditions constant across sites.

\section*{Discussion}
The tension between the benchmark and operational models exposes a gap between what replication practice implicitly assumes and what it actually does. To reiterate, the data structure of typical replication practice resembles the operational model (with $m > 1,$ all $k = 1$). This is the regime in which the benchmark and operational models are observationally indistinguishable, while metascience's inferential practice implicitly invokes the benchmark's interpretation of a shared replicability rate when reporting and comparing rates. The forms of misuse we discuss in the demarcation, aggregation, power comparison, reference, and crisis subsections below all trace back to this fundamental mismatch between the question being asked and the model being used to answer it.

\subsection*{The demarcation problem}
Using replications of individual results to demarcate between those that hold up and those that do not is a common interpretive practice. We showed that even when the result type is fixed across replications, making useful decisions on replicability based on dichotomous decisions is not possible under the realistic condition of non-exact replications, regardless of how many replications are performed. Further, reported replicability rates are compared across journals, disciplines, and time in current metascientific practice~\citep[e.g.,][]{open2015estimating,tyner2026investigating}, with the aim of distinguishing healthy from unhealthy literatures. Such comparisons lack a sound statistical foundation, for the same reasons that apply to individual results. Assuming $\rho = 0$ does not eliminate the between-experiment heterogeneity that drives these consequences. \textit{The distinction between results that hold and results that do not cannot be reliably drawn from replicability rates produced by current methods.}

\subsection*{The aggregation problem}
Heterogeneity across results, which entails the variability between distinct results within a research area or between research areas, invokes at least a hierarchical structure that current practice also fails to address. Literatures delineated by areas of study can be modeled to have a distribution of replicability rates of their results. Mature fields may expect a high frequency of high-replicability results, whereas emerging or stalled fields may expect a lower frequency. Some results may be studied under near-exact replication regimes, while others are investigated under substantial heterogeneity. Typical metascientific practice of referring to an average replicability rate for a diverse population of results~\citep[e.g.,][]{camerer2016evaluating,open2015estimating}, sometimes even across disciplines~\citep[e.g.,][]{tyner2026investigating}, is statistically fraught and misleading. An average that suppresses a joint distribution does not provide a valid summary of the aggregated data. 

The limitation of such ad hoc estimates becomes clear when we consider the replicability rate in a literature as arising from a mixture distribution. A class of results, defined by its underlying phenomenon, theories and models, replication regime, and methodological conventions, contributes its own $(\mu_k, \rho_k)$ to the replicability profile of the literature. A mathematically valid estimated literature-wide rate of the mixture mean $\sum_k w_k \mu_k,$ where $w_k$ are mixing probabilities, can be obtained. This estimate, however, is not interpretable because it conflates the mean of a heterogeneous distribution with a property of individual results that differ fundamentally in their scientific character. The constituent classes likely differ in $\rho_k,$ so the mixture mean reduces quantities of different precision into a single number whose variance is a weighted combination of within-class variances $\mu_k(1-\mu_k)[1+(m_k-1) \rho_k]/m_k.$ These variances can differ by orders of magnitude across classes. Without modeling the distributions of sampled literatures, the existence of an ``expected replicability rate'' against which the estimated rate can be compared remains an assumption that is never stated, never justified, and never testable from the replication data alone. {\em Matters do not improve when replicability rates are produced by aggregating heterogeneous literatures.}

\subsection*{The power comparison problem}
Comparing the estimated mean replicability rate to the estimated mean power of replication studies is a common practice in large-scale replication projects. The implicit claim is that the two quantities should agree if the scientific process is operating correctly. Typically, large-scale replication studies are statistically powered to detect the effect size reported in a reference and they note whether it is different from the rate of successful replications~\citep[e.g.,][]{open2015estimating,tyner2026investigating}. Any discrepancy is then treated as informative about the state of the literature. Such comparisons are ill-founded.

In frequentist statistics, statistical power is defined for a specific test as a function of sample size and effect size. It is not a property of a result in the abstract. The mean replicability rate, by contrast, is defined for any statistically conceivable result type, $R,$ (e.g., a confidence interval, a Bayes factor, a probability threshold). Even when $R$ is consistently defined as the outcome of a null hypothesis significance test across all labs, the comparison of outcomes requires that the definition of $R$ is held constant across power calculations, which is not necessarily the case in practice. Beyond the definitional mismatch, reference studies are selected as targets in replication projects because they produced significant results, so effect size estimates are biased upward by selection. This phenomenon is sometimes referred to as the winner's curse~\citep{button2013power}, or can be more generally considered a type~M error~\citep{gelman2014beyond}. Ultimately it inflates the power benchmark before the comparison begins. Averaging power across non-exact replications using the reference effect size as a plug-in is equivalent to computing $\mu$ under exact replications. The comparison of statistical power to estimated replicability rate neither validates a replication methodology nor diagnoses its failures. {\em Comparing a quantity computed under exact replication, and known and fixed $R$ to an estimate whose variance is dominated by between-lab heterogeneity the comparison ignores is an invalid statistical practice.}

\subsection*{The reference problem}
A natural solution to the discriminability problem is to strive for exact replications. If non-exactness can be driven close to zero by tightening design controls, the asymptotic floor on the variance of replicability rates decreases, and meaningful inference becomes possible in principle. The ML4 analysis raises a practical concern however. The AA protocol received direct procedural guidance from the original authors, with standardized scripts, materials, and experimenter instructions. Yet the quantitative evaluation shows that even within the AA group, design standardization did not suffice to eliminate non-exactness. It may have reduced some sources of heterogeneity but could not overcome the amplification of between-site differences that arises from applying a significance test to near-zero effects at finite sample sizes. The reason is that controlling design cannot control all sources of variability in effect sizes or replicability rates. Sampling variability, measurement error, and context sensitivity contribute to $\rho$ in ways that no protocol can fully suppress without a clear understanding of what generates the effect in the first place. This brings us to a second difficulty. The replication sequence is typically anchored to a reference experiment that itself is noisy. Conditioning replication sequences on any other study equals treating the reference as the fixed point toward which replications converge, injecting non-exactness in any replication sequence. The reference experiment should not be treated as a gold standard unless such a privileged position can be justified on its own merits.

If a replication sequence need not converge on the design of the reference experiment, what, then, should it be anchored on? Two alternatives can be considered. The first is a theoretical prediction. If a strong theory specifies the effect, its boundary conditions, and the population to which it applies, the most precise study design constrained by that prediction becomes the natural anchor. Underspecified theories yield underdetermined anchors, increasing arbitrariness. The second alternative is an elementary empirical anchor. That is the logic behind the minimum viable experiment framework~\citep{devezer2025mve}, which defines the minimal set of experimental parameters required to generate an empirical regularity. Here, dependence on theory is minimized and the focus falls on identifying the indispensable conditions to produce the result. Replication sequences organized around such a minimal form, rather than an arbitrary reference experiment, can in principle achieve small $\rho$ by eliminating all auxiliary sources of variability. Genuine progress on replicability therefore requires either advancing theory or systematically eliminating empirical noise. \textit{Design standardization around an arbitrary reference experiment can substitute for neither a strong theoretical account of what is being replicated, nor a precise experimental design that is free of all but sampling variability.}

\subsection*{The crisis problem}
Implications of our framework extend to the broader narrative of replication crisis. Declarations of crisis rest on the implicit inferential claim that estimated replicability rates~\citep[such as the $36\%$ reported in the RPP,][]{open2015estimating} are meaningfully lower than what a highly replicable literature would produce. This position is statistically untenable for two reasons. 

The first is the aggregation problem already established above. The $36\%$ rate pools results from different fields, different effect sizes, and different replication conditions. Whatever its statistical precision, it estimates a quantity with no stable scientific interpretation. A precise estimate of an uninterpretable quantity is not evidence of a crisis. The second concerns what the $36\%$ rate could establish even if the aggregation problem did not exist. For a fixed result, declaring a crisis requires showing that the estimated rate is genuinely low. But low relative to what? Without specifying the base rate of genuine effects in the sampled literature, there is no expected replicability rate against which to compare the estimated one. A rate of $36\%$ could be consistent with a healthy literature full of small, noisy studies or a literature riddled with false positives. Replication data do not let us distinguish these cases, both because $\rho$ is non-identifiable from binary verdicts and because the base rate is unspecified.

A crisis label requires the ability to read signal from the data. The signal is not there to read with current replication methods. \textit{The crisis, if it exists, cannot be established by the methods used to declare it.\footnote{Accounts of the events that catalyzed the crisis narrative vary, and some prominent fraud cases and methodological critiques predate the large-scale replication projects. Our claim does not depend on this chronology. Isolated replication failures, however prominent, do not license the label ``crisis''. Science has always contained failed replications and fraud cases do not specifically license a ``replication'' crisis per se. The label ``replication crisis'' asserts that failures are systematic rather than idiosyncratic, and a claim of systematic failure is a claim about the replicability rate of a literature, whether or not a rate has been computed. Large-scale replicability studies~\citep[e.g.,][]{open2015estimating} formalized this claim~\citep[see also][]{sep-scientific-reproducibility} even if they may not have created it. It is this claim, in its informal and formal versions alike, that our analysis shows cannot be established from binary verdict data.}}

\subsection*{What replication is good for}

We would like to caution against a potential misinterpretation of our argument. We do not speak against the general utility of performing replications. We argue against a specific use of replication outcomes, in particular their reduction to binary verdicts and treatment as a demarcation criterion. Replication remains valuable for estimating effect magnitudes, characterizing variability across contexts, calibrating measurements, finetuning experimental procedures, and informing cumulative inference. None of these aims requires the discriminability of high- and low-replicability sequences. We do not claim that meaningful replication is impossible. We claim that the specific data structure and analytical methods currently used to declare replicability rates and pronounce verdicts on fields are inadequate for those tasks.

The conversion of a replicability rate, computed from coarsened data, into a verdict on individual results, on fields, or on science is unwarranted. Replication has served science well for centuries and will continue to serve it in many forms and capacities, but not as a tribunal.

\subsection*{Acknowledgements}
\noindent We thank Andrew Gelman, Erik van Zwet, Dean Shamess, and Alex Hayes for thoughtful comments on a previous draft of this work.

\section*{Appendix} \label{sec:appendix}
\begin{enumerate}

\item {\bf Variance of $\hat{\mu}$ under the operational model.}

Since the $X_i$ are independent across experiments, we have
$$\mathbb{V}(\hat{\mu} \mid k_i, m, \mu, \rho) = \frac{1}{m^2}
\sum_{i=1}^{m} \mathbb{V}(X_i/k_i \mid k_i, \mu, \rho).$$
For each experiment $i,$ the marginal variance of $X_i/k_i$ under the Betabinomial is
$$\mathbb{V}(X_i/k_i \mid k_i, \mu, \rho) =
\frac{\mu(1-\mu)}{k_i}\left[1 + (k_i - 1)\rho\right]
= \mu(1-\mu)\left[\frac{1-\rho}{k_i} + \rho\right].$$
Substituting and separating terms that depend on $k_i$ from those that do not,
$$\mathbb{V}(\hat{\mu} \mid k_i, m, \mu, \rho) = \mu(1-\mu)\left[\frac{1-\rho}{m^2}
\sum_{i=1}^{m}\frac{1}{k_i}\;+\; \frac{\rho}{m}\right].$$
The two limiting cases follow directly.
\begin{itemize}
    \item When all $k_i \to \infty,$ the terms $(1-\rho)/k_i$ vanish and
    $$\mathbb{V}(\hat{\mu} \mid m, \mu, \rho) = \frac{\mu(1-\mu)\rho}{m}.$$
    \item When all $k_i = 1,$ we have $\sum_{i=1}^m(1/k_i) = m,$ so
    $$\mathbb{V}(\hat{\mu} \mid m, \mu)
    = \mu(1-\mu) \left[\frac{1-\rho}{m} + \frac{\rho}{m}\right]
    = \frac{\mu(1-\mu)}{m},$$
and $\rho$ drops out.
\end{itemize}

\item {\bf Excess variance of the operational model at finite $k$ relative to the $k \to \infty$ limit.}

\begin{eqnarray*}
\mathbb{V}(\hat{\mu}|k, m, \mu, \rho) - \mathbb{V}(\hat{\mu}|k \to \infty, m, \mu, \rho) 
&=&\frac{\mu(1-\mu)}{mk}[1+(k-1)\rho]-\frac{\mu(1-\mu)\rho}{m}\\
&=&\frac{\mu(1-\mu)(1-\rho)}{mk}. 
\end{eqnarray*}

\item {\bf The mean replicability rate in the first example: Effect of sample heterogeneity on replicability rate.}

Let $Z \sim \mathrm {Nor}(0,1).$ We have
$$\mu = \mathbb{E}_{\varepsilon_i}\!\left[\Phi\!\left(
\frac{\theta + \varepsilon_i}{\mathrm{SE}}\right)\right] =\mathbb{E}\left[ P\left(Z\leq \frac{\theta+\varepsilon_i}{SE}|\varepsilon_i\right)\right] = 
P\!\left(Z \cdot \mathrm{SE} - \varepsilon_i \leq \theta\right).$$
$Z \cdot \mathrm{SE} \sim \mathrm {Nor}(0, \mathrm{SE}^2)$ and
$\varepsilon_i \sim \mathrm {Nor}(0, \sigma^2)$ are independent,
implying $(Z\cdot SE - \varepsilon) \sim \mathrm {Nor}(0, \mathrm{SE}^2 + \sigma^2).$ We have
$$\mu = P\!\left(Z \cdot \mathrm{SE} - \varepsilon_i \leq
\theta\right) = \Phi\!\left(\frac{\theta}{\sqrt{\mathrm{SE}^2 +
\sigma^2}}\right).$$

\item {\bf The intraclass correlation parameter in the first
example: Effect of sample heterogeneity on replicability rate.}
We have $\phi_i = \Phi((\theta + \varepsilon_i)/\mathrm{SE})$ where
$\varepsilon_i \sim \mathrm {Nor}(0, \sigma^2)$ are iid across experiments. Hence, $\phi_i$ are iid with mean 
$\mathbb{E}[\phi_i] = \mu.$ We write the variance of $\phi_i$ as
\begin{equation}\label{eq:shorthand}
\mathbb{V}(\phi_i) = \mathbb{E}[\phi_i^2] - \left(\mathbb{E}[\phi_i]\right)^2
= \mathbb{E}[\phi_i^2] - \mu^2.
\end{equation}
By definition, the intraclass correlation $\rho$ is 
$$\rho = \frac{\mathbb{V}(\phi_i)}{\mu(1-\mu)}.$$
When $\phi_i$ is
degenerate (all $\phi_i = \mu,$ corresponding to exact
replication), $\mathbb{V}(\phi_i) = 0$ and $\rho = 0.$ When
$\phi_i \in \{0,1\},$ binary verdict, we have
$\mathbb{V}(\phi_i) = \mu(1-\mu)$ and $\rho = 1.$ Substituting~\ref{eq:shorthand}, we get
\begin{equation*}
\rho = \frac{\mathbb{E}[\phi_i^2] - \mu^2}{\mu(1-\mu)}
= \frac{\mathbb{E}\!\left[\Phi\!\left(\frac{\theta +
\varepsilon_i}{\mathrm{SE}}\right)^2\right] - \mu^2}{\mu(1-\mu)}.
\end{equation*}
To compute $\mathbb{E}[\phi_i^2],$ we write
$$\mathbb{E}[\phi_i^2] = \mathbb{E}_{\varepsilon_i}\!\left[
\Phi\!\left(\frac{\theta + \varepsilon_i}{\mathrm{SE}}
\right)^2\right] = \int_{-\infty}^{\infty}
\Phi\!\left(\frac{\theta + \varepsilon}{\mathrm{SE}}\right)^2
\frac{1}{\sigma}\varphi\!\left(\frac{\varepsilon}{\sigma}\right)
d\varepsilon,$$
where $\varphi$ is the Standard Normal density. This integral
has no closed form and is evaluated numerically. 

We write
$$\Phi((\theta + \varepsilon)/\mathrm{SE})^2 =
P(Z_1 \leq (\theta + \varepsilon)/\mathrm{SE})\cdot
P(Z_2 \leq (\theta + \varepsilon)/\mathrm{SE}),$$
where $Z_1, Z_2$ are independent standard Normal random
variables. Then
\begin{align*}
\mathbb{E}[\phi_i^2] &= P\!\left(Z_1 \cdot \mathrm{SE}
\leq \theta + \varepsilon,\;
Z_2 \cdot \mathrm{SE} \leq \theta + \varepsilon\right) \\
&= P\!\left(Z_1 \cdot \mathrm{SE} - \varepsilon \leq \theta,\;
Z_2 \cdot \mathrm{SE} - \varepsilon \leq \theta\right).
\end{align*}
The joint distribution of $(Z_1 \cdot \mathrm{SE} - \varepsilon,\;
Z_2 \cdot \mathrm{SE} - \varepsilon)$ is bivariate Normal with
mean $(0, 0)$ and covariance matrix
$$\Sigma = \begin{pmatrix} \mathrm{SE}^2 + \sigma^2 & \sigma^2 \\
\sigma^2 & \mathrm{SE}^2 + \sigma^2 \end{pmatrix},$$
because the two components share the same $\varepsilon$ term,
inducing correlation $\sigma^2/(\mathrm{SE}^2 + \sigma^2).$
Therefore
$$\mathbb{E}[\phi_i^2] = \Phi_2\!\left(\frac{\theta}
{\sqrt{\mathrm{SE}^2 + \sigma^2}},\,
\frac{\theta}{\sqrt{\mathrm{SE}^2 + \sigma^2}};\,
\frac{\sigma^2}{\mathrm{SE}^2 + \sigma^2}\right),$$
where $\Phi_2(\cdot, \cdot\,; r)$ is the standard bivariate
Normal CDF with correlation $r.$ This expression is evaluated
numerically for each $(\theta, \sigma, \mathrm{SE})$ combination
reported in Table~\ref{tab:composition}.

In the large-$n$ limit where $\mathrm{SE} \to 0,$ we have 
$$\mathbb{E}[\phi_i^2] \to \Phi_2\!\left(\frac{\theta}{\sigma},\,
\frac{\theta}{\sigma};\, 1\right) =
\Phi\!\left(\frac{\theta}{\sigma}\right),$$
since correlation $1$ collapses the bivariate Normal to a
univariate one. The intraclass correlation then becomes
$$\rho = \frac{\Phi(\theta/\sigma) - \mu^2}{\mu(1-\mu)}.$$
With $\mu = \Phi(\theta/\sigma)$ in this limit, we have
$\rho = (\mu - \mu^2)/(\mu(1-\mu)) = 1,$ which is the
degenerate case where all $\phi_i$ collapse to $0$ or $1.$
For large but finite $n,$ the numerical evaluation of
$\Phi_2$ gives the values reported in
Table~\ref{tab:composition}.

\item {\bf Many Labs 4 priors.}

When $g_i\sim \mathrm {Nor}(\theta, \sigma^2)$ are independent and identically distributed, conjugate priors are $$\theta| \sigma^2 \sim \mathrm {Nor}(\mu_0, \sigma^2/\kappa_0),\; \sigma^2 \sim \text{Inverse-Gamma}(\alpha_0, \beta_0).$$ Posterior parameter updates under conjugacy are given by  $$\kappa_n = \kappa_0 + n, \quad \mu_n = (\kappa_0 \mu_0 + n\bar{g})/\kappa_n, \quad \alpha_n = \alpha_0 + n/2,$$
$$\beta_n = \beta_0 + \frac{1}{2}\sum_{i=1}^n(g_i - \bar{g})^2 + [\kappa_0 n(\bar{g}-\mu_0)^2]/(2\kappa_n),$$  where $\bar{g}$ is the mean of $18$ $g_i.$ The marginal posterior distribution of $\theta$ is Student's $t$, and of $\sigma^2$ is Inverse-Gamma. We sample 300,000 values of $\theta$ and  $\sigma^2$ from the Normal-Inverse-Gamma posteriors directly. 

For hyperparameter choices, we use two sets. First we use Jeffreys priors 
$\kappa_0 \to 0,\; \alpha_0 \to 0,\; \beta_0 \to 0,$ which is improper and gives
$P(\theta, \sigma^2) \propto 1/\sigma^3$  but the posterior is proper (for $n\geq 2$). This is the standard non-informative prior since it implies that $\theta$ is Uniform on the real line, carrying minimal information about where the mean effect sits. The marginal prior on $\sigma^2$ is proportional to $1/\sigma^2.$ The posterior for $\theta$ is Student-t centered at $\bar{g}$ with scale $s_g/\sqrt{n}$ and $n$ degrees of freedom, the sampling distribution of the mean. The posterior for $\sigma^2$ is an Inverse-Gamma centered near $s_g^2.$ 

When $\kappa_0 \to 0$ in the Normal-Inverse-Gamma prior, the variance $\sigma^2/\kappa_0 \to \infty$ so the Normal flattens to a constant. It becomes proportional to 1 for any finite $\theta$. The conditional $\theta|\sigma^2 \propto 1$ in this limit, which is the Uniform improper prior on the real line. The Normal-Inverse-Gamma family recovers the Uniform prior on $\theta$ in the limit, and for any finite $\kappa_0 > 0$ the conditional is a proper Normal distribution.

The conjugate Normal conditional prior on $\theta | \sigma^2$ and the flat prior on $\theta$ are the same distribution in the limit $\kappa_0 \to 0.$ They are not two different priors since the flat prior is the degenerate limiting member of the Normal family as the variance goes to infinity. The construction is standard: improper priors are often obtained as limiting cases of proper conjugate priors. For finite $\kappa_0,$ the marginal distribution of $\theta$ is a Student's t with heavy-tails because it integrates over uncertainty in $\sigma^2.$ In the limit $\kappa_0 \to 0, \;\alpha_0 \to 0,\; \beta_0 \to 0,$ the Student's t also flattens to a constant, recovering $p(\theta) \propto 1.$

The key distinction between Jeffreys and the weakly informative prior is that $\beta_0 = 1$ in the weakly informative case places substantial prior mass on moderate to large values of $\sigma^2$ relative to the actual spread of the ML4 $g$ values, which is why $\rho$ is inflated under the weakly informative prior, particularly for the small $m$ cases where the prior has more leverage.

\item {\bf Finite-$n$ correction for the ML4 posterior analysis in Tables~\ref{tab:ML4} and ~\ref{tab:AAIH}.}

The auxiliary Normal model fits $g_i \sim \mathrm {Nor}(\theta, \sigma^2)$
and maps posterior draws of $(\theta, \sigma^2)$ to $(\mu, \rho)$ via
Monte Carlo simulation. The large-$n$ mapping $\phi_i = \Phi(\theta_i)$ treats the standard error of each effect size estimate as $1,$ which corresponds to the limit where within-experiment sampling variability is negligible. For the ML4 studies, within-experiment sampling variability is not negligible given the sample sizes involved.

The finite-$n$ correction replaces the unit standard error with the actual standard error of Hedges' $g$ at each site. For two independent groups of sizes $n_{1i}$ and $n_{2i},$ the standard error is
\begin{equation*}
\mathrm{SE}_i = \sqrt{\frac{n_{1i}+n_{2i}}{n_{1i}\,n_{2i}}
+ \frac{g_i^2}{2(n_{1i}+n_{2i}-2)}}.
\end{equation*}
The corrected mapping is $\phi_i = \Phi(\theta_i / \mathrm{SE}_i),$ which is the probability that a directional test rejects when $(\theta_i / \mathrm{SE}_i)>0.$ Distributions for parameters $(\mu,\rho)$ are obtained using Monte Carlo integration with $300,000$ draws, using each site specific $\mathrm{SE}_i.$

\end{enumerate}

\bibliography{Biblio}

@article{begley2012raise,
  title={Raise standards for preclinical cancer research},
  author={Begley, C Glenn and Ellis, Lee M},
  journal={Nature},
  volume={483},
  number={7391},
  pages={531--533},
  year={2012},
  publisher={Nature Publishing Group UK London}
}

@article{bryan2021behavioural,
  title={Behavioural science is unlikely to change the world without a heterogeneity revolution},
  author={Bryan, Christopher J and Tipton, Elizabeth and Yeager, David S},
  journal={Nature human behaviour},
  volume={5},
  number={8},
  pages={980--989},
  year={2021},
  publisher={Nature Publishing Group UK London}
}

@article{button2013power,
  title={Power failure: why small sample size undermines the reliability of neuroscience},
  author={Button, Katherine S and Ioannidis, John PA and Mokrysz, Claire and Nosek, Brian A and Flint, Jonathan and Robinson, Emma SJ and Munaf{\`o}, Marcus R},
  journal={Nature reviews neuroscience},
  volume={14},
  number={5},
  pages={365--376},
  year={2013},
  publisher={Nature Publishing Group UK London}
}

@article{buzbas2023,
  title={The logical structure of experiments lays the foundation for a theory of reproducibility},
  author={Buzbas, Erkan O and Devezer, Berna and Baumgaertner, Bert},
  journal={Royal Society Open Science},
  volume={10},
  number={3},
  pages={221042},
  year={2023},
  publisher={The Royal Society}
}

@article{camerer2016evaluating,
  title={Evaluating replicability of laboratory experiments in economics},
  author={Camerer, Colin F and Dreber, Anna and Forsell, Eskil and Ho, Teck-Hua and Huber, J{\"u}rgen and Johannesson, Magnus and Kirchler, Michael and Almenberg, Johan and Altmejd, Adam and Chan, Taizan and others},
  journal={Science},
  volume={351},
  number={6280},
  pages={1433--1436},
  year={2016},
  publisher={American Association for the Advancement of Science}
}

@article{open2015estimating,
  title={Estimating the reproducibility of psychological science},
  author={Open Science Collaboration},
  journal={Science},
  volume={349},
  number={6251},
  pages={aac4716},
  year={2015},
  publisher={American Association for the Advancement of Science}
}

@article{baker2016reproducibility,
  title={Reproducibility crisis},
  author={Baker, Monya},
  journal={Nature},
  volume={533},
  number={26},
  pages={353--66},
  year={2016}
}

@article{Pashler2012,
author = {Harold Pashler and Eric–Jan Wagenmakers},
title ={Editors’ Introduction to the Special Section on Replicability in Psychological Science: A Crisis of Confidence?},
journal = {Perspectives on Psychological Science},
volume = {7},
number = {6},
pages = {528-530},
year = {2012},
doi = {10.1177/1745691612465253},
    note ={PMID: 26168108},
URL = {https://doi.org/10.1177/1745691612465253},
eprint = {https://doi.org/10.1177/1745691612465253}
}

@article{freedman2015economics,
  title={The economics of reproducibility in preclinical research},
  author={Freedman, Leonard P and Cockburn, Iain M and Simcoe, Timothy S},
  journal={PLoS biology},
  volume={13},
  number={6},
  pages={e1002165},
  year={2015},
  publisher={Public Library of Science}
}

@article{errington2021challenges,
  title={Challenges for assessing replicability in preclinical cancer biology},
  author={Errington, Timothy M and Denis, Alexandria and Perfito, Nicole and Iorns, Elizabeth and Nosek, Brian A},
  journal={elife},
  volume={10},
  pages={e67995},
  year={2021},
  publisher={eLife Sciences Publications, Ltd}
}

@InCollection{sep-scientific-reproducibility,
	author       =	{Fidler, Fiona and Wilcox, John},
	title        =	{{Reproducibility of Scientific Results}},
	booktitle    =	{The {Stanford} Encyclopedia of Philosophy},
	editor       =	{Edward N. Zalta and Uri Nodelman},
	howpublished =	{\url{https://plato.stanford.edu/archives/sum2026/entries/scientific-reproducibility/}},
	year         =	{2026},
	edition      =	{{S}ummer 2026},
	publisher    =	{Metaphysics Research Lab, Stanford University}
}

@article{devezer2025mve,
    title = {Minimum Viable Experiment and the Illusion of Diagnostic Replication},
    author = {Berna Devezer and Erkan O Buzbas},
    journal = {Minnesota Studies in the Philosophy of Science},
    year = {in press}
}

@article{gelman2014beyond,
  title={Beyond power calculations: Assessing type S (sign) and type M (magnitude) errors},
  author={Gelman, Andrew and Carlin, John},
  journal={Perspectives on psychological science},
  volume={9},
  number={6},
  pages={641--651},
  year={2014},
  publisher={Sage Publications Sage CA: Los Angeles, CA}
}

@article{gelman2006difference,
  title={The difference between “significant” and “not significant” is not itself statistically significant},
  author={Gelman, Andrew and Stern, Hal},
  journal={The American Statistician},
  volume={60},
  number={4},
  pages={328--331},
  year={2006},
  publisher={Taylor \& Francis}
}

@article{greenberg1994role,
  title={Role of consciousness and accessibility of death-related thoughts in mortality salience effects.},
  author={Greenberg, Jeff and Pyszczynski, Tom and Solomon, Sheldon and Simon, Linda and Breus, Michael},
  journal={Journal of personality and social psychology},
  volume={67},
  number={4},
  pages={627},
  year={1994},
  publisher={American Psychological Association}
}

@article{hedges1981distribution,
  title={Distribution theory for Glass's estimator of effect size and related estimators},
  author={Hedges, Larry V},
  journal={journal of Educational Statistics},
  volume={6},
  number={2},
  pages={107--128},
  year={1981},
  publisher={Sage Publications Sage CA: Thousand Oaks, CA}
}

@article{henrich2010most,
  title={Most people are not WEIRD},
  author={Henrich, Joseph and Heine, Steven J and Norenzayan, Ara},
  journal={Nature},
  volume={466},
  number={7302},
  pages={29--29},
  year={2010},
  publisher={Nature Publishing Group UK London}
}

@article{klein2022many,
  title={Many Labs 4: Failure to replicate mortality salience effect with and without original author involvement},
  author={Klein, Richard A and Cook, Corey L and Ebersole, Charles R and Vitiello, Christine and Nosek, Brian A and Hilgard, Joseph and Ahn, Paul Hangsan and Brady, Abbie J and Chartier, Christopher R and Christopherson, Cody D and others},
  journal={Collabra: Psychology},
  volume={8},
  number={1},
  pages={35271},
  year={2022},
  publisher={University of California Press}
}

@article{klein2018many,
  title={Many Labs 2: Investigating variation in replicability across samples and settings},
  author={Klein, Richard A and Vianello, Michelangelo and Hasselman, Fred and Adams, Byron G and Adams Jr, Reginald B and Alper, Sinan and Aveyard, Mark and Axt, Jordan R and Babalola, Mayowa T and Bahn{\'\i}k, {\v{S}}t{\v{e}}p{\'a}n and others},
  journal={Advances in Methods and Practices in Psychological Science},
  volume={1},
  number={4},
  pages={443--490},
  year={2018},
  publisher={Sage Publications Sage CA: Los Angeles, CA}
}

@article{meng2020reproducibility,
  title={Reproducibility, replicability, and reliability},
  author={Meng, Xiao-Li},
  journal={Harvard Data Science Review},
  volume={2},
  number={4},
  pages={10},
  year={2020},
  publisher={The MIT Press}
}

@article{niepel2019multi,
  title={A multi-center study on the reproducibility of drug-response assays in mammalian cell lines},
  author={Niepel, Mario and Hafner, Marc and Mills, Caitlin E and Subramanian, Kartik and Williams, Elizabeth H and Chung, Mirra and Gaudio, Benjamin and Barrette, Anne Marie and Stern, Alan D and Hu, Bin and others},
  journal={Cell systems},
  volume={9},
  number={1},
  pages={35--48},
  year={2019},
  publisher={Elsevier}
}

@article{prinz2011believe,
  title={Believe it or not: how much can we rely on published data on potential drug targets?},
  author={Prinz, Florian and Schlange, Thomas and Asadullah, Khusru},
  journal={Nature Reviews Drug Discovery},
  volume={10},
  number={9},
  pages={712--712},
  year={2011},
  publisher={Nature Publishing Group UK London}
}

@article{tyner2026investigating,
  title={Investigating the replicability of the social and behavioural sciences},
  author={Tyner, Andrew H and Abatayo, Anna Lou and Daley, Mason and Field, Samuel and Fox, Nicholas and Haber, Noah A and Hahn, Krystal M and Struhl, Melissa Kline and Mawhinney, Brinna and Miske, Olivia and others},
  journal={Nature},
  volume={652},
  number={8108},
  pages={143--150},
  year={2026},
  publisher={Nature Publishing Group UK London}
}

@article{ward2015achieving,
  title={Achieving across-laboratory replicability in psychophysical scaling},
  author={Ward, Lawrence M and Baumann, Michael and Moffat, Graeme and Roberts, Larry E and Mori, Shuji and Rutledge-Taylor, Matthew and West, Robert L},
  journal={Frontiers in Psychology},
  volume={6},
  pages={903},
  year={2015},
  publisher={Frontiers Media SA}
}

@article{watson1988development,
  title={Development and validation of brief measures of positive and negative affect: the PANAS scales.},
  author={Watson, David and Clark, Lee Anna and Tellegen, Auke},
  journal={Journal of personality and social psychology},
  volume={54},
  number={6},
  pages={1063},
  year={1988},
  publisher={American Psychological Association}
}

\onecolumn
\newpage

\end{document}